\documentclass[%
 reprint,
superscriptaddress,
showpacs,
 amsmath,
 amssymb,
 amsfonts,
 aps,
 pra,
showkeys
]{revtex4-1}

\RequirePackage{atbegshi}

\usepackage{dcolumn}        
\usepackage{bbold}          
\usepackage{verbatim}       
\usepackage[dvipsnames, table]{xcolor}
\usepackage{xparse}
\usepackage{siunitx}
\usepackage{subfigure}      
\renewcommand{\thesubfigure}{\alph{subfigure}}
\makeatletter
  \renewcommand{\@thesubfigure}{(\thesubfigure)\space}
  \def\@currentlabel{\p@subfigure\thesubfigure}
\makeatother

\usepackage{units}          
\usepackage[]{todonotes}    
\usepackage[matha]{mathabx} 
\usepackage{tabularx}
\usepackage{multirow}
\usepackage{booktabs}       

\usepackage[utf8]{inputenc}
\usepackage[T1]{fontenc}
\usepackage[ngerman,english]{babel}
\usepackage{slashed}
\usepackage{upgreek}
\usepackage{dsfont}

\usepackage{hyperref}
\hypersetup{
    colorlinks=true,
    citecolor=black,
    linkcolor=black,
    urlcolor=black,
    anchorcolor=black,
    linktocpage
}

\newcommand{\Tr}{\operatorname{Tr}}
\DeclareMathOperator{\sign}{sgn}
\newcommand{\Order}{\ensuremath{\mathcal{O}}}

\newcommand{\NSigma}{\ensuremath{\text{N}_\sigma}}
\newcommand{\NTau}{\ensuremath{\text{N}_\tau}}
\newcommand{\Nc}{\ensuremath{\text{N}_\text{c}}}
\newcommand{\Nf}{\ensuremath{\text{N}_\text{f}}}

\newcommand{\Loewe}{LOEWE-CSC}
\newcommand{\Lcsc}{L-CSC}
\newcommand{\Qcd}{QCD}
\newcommand{\clqcd}{CL\kern-.25em\textsuperscript{2}QCD}
\newcommand{\Ocl}{OpenCL}
\newcommand{\referencename}{Ref.}
\newcommand{\sectionname}{Sec.}
\newcommand{\Eq}[1]{Eq.~\eqref{#1}}
\newcommand{\Eqs}[1]{Eqs.~\eqref{#1}}

\newcommand{\psibar}{\bar{\psi}} 
\newcommand{\chiralcond}{\ensuremath{\langle \psibar \psi \rangle}}
\newcommand{\mpi}{\ensuremath{m_{\pi}}}
\newcommand{\PartFunc}{\ensuremath{\mathcal Z}}
\newcommand{\Action}{\ensuremath{\mathcal S}}
\newcommand{\SGluon}{\ensuremath{\Action_{\text{g}}}}

\newcommand{\Link}{\ensuremath{U}}
\newcommand{\LatSpacing}{\ensuremath{a}}
\newcommand{\LatMassWilson}{\ensuremath{\kappa}}
\newcommand{\LatMassWilsonTric}{\ensuremath{\LatMassWilson^\text{tricr.}}}
\newcommand{\LatMassWilsonTricHeavy}{\ensuremath{\LatMassWilsonTric_\text{heavy}}}
\newcommand{\LatMassWilsonTricLight}{\ensuremath{\LatMassWilsonTric_\text{light}}}
\newcommand{\LatCoupling}{\ensuremath{\beta}}
\newcommand{\LatCouplingC}{\ensuremath{\LatCoupling_c}}
\newcommand{\PL}{Polyakov loop}
\newcommand{\Poly}{\ensuremath{L}}
\newcommand{\PolyIm}{\ensuremath{\Poly_\text{Im}}}
\newcommand{\PolyRe}{\ensuremath{\Poly_\text{Re}}}

\newcommand{\Binder}{\ensuremath{B_4}}
\newcommand{\Temp}{\ensuremath{T}}
\newcommand{\Tc}{\ensuremath{\Temp_c}}
\newcommand{\MuI}{\ensuremath{\mu_i}}
\newcommand{\RW}{Roberge-Weiss}
\newcommand{\TauInt}{\ensuremath{\tau_\text{int}}}

\begin{document}

\title{Roberge-Weiss transition in \texorpdfstring{$\Nf=2$}{Nf=2} QCD with Wilson
       fermions and \texorpdfstring{$\NTau=6$}{Nt=6}}

\author{Christopher Czaban}
 \email{czaban@th.physik.uni-frankfurt.de}
 \affiliation{
 Institut f\"{u}r Theoretische Physik - Johann Wolfgang Goethe-Universit\"{a}t, Germany\\
 Max-von-Laue-Str.\ 1, 60438 Frankfurt am Main
}
\affiliation{
John von Neumann Institute for Computing (NIC)
GSI, Planckstr.\ 1, 64291 Darmstadt, Germany
}

\author{Francesca Cuteri}
 \email{cuteri@th.physik.uni-frankfurt.de}
\affiliation{
 Institut f\"{u}r Theoretische Physik - Johann Wolfgang Goethe-Universit\"{a}t, Germany\\
 Max-von-Laue-Str.\ 1, 60438 Frankfurt am Main
}

\author{Owe Philipsen}
 \email{philipsen@th.physik.uni-frankfurt.de}
\affiliation{
 Institut f\"{u}r Theoretische Physik - Johann Wolfgang Goethe-Universit\"{a}t, Germany\\
 Max-von-Laue-Str.\ 1, 60438 Frankfurt am Main
}
\affiliation{
John von Neumann Institute for Computing (NIC)
GSI, Planckstr.\ 1, 64291 Darmstadt, Germany
}

\author{Christopher Pinke}
 \email{pinke@th.physik.uni-frankfurt.de}

\author{Alessandro Sciarra}
 \email{sciarra@th.physik.uni-frankfurt.de}
\affiliation{
 Institut f\"{u}r Theoretische Physik - Johann Wolfgang Goethe-Universit\"{a}t, Germany\\
 Max-von-Laue-Str.\ 1, 60438 Frankfurt am Main
}

\date{December 22, 2015}

\begin{abstract}
 QCD with imaginary chemical potential is free of the sign problem and exhibits a rich phase structure 
  constraining the phase diagram at real chemical potential. 
  We simulate the critical endpoint of the \RW{} (RW) transition at imaginary chemical 
  potential for $\Nf=2$ QCD on $\NTau=6$ lattices with standard Wilson fermions.
  As found on coarser lattices, the RW endpoint is a triple point connecting the deconfinement/chiral transitions in the heavy/light quark mass region and changes to a second-order endpoint for intermediate masses.
  These regimes are separated by two tricritical values of the quark mass,
    which we determine by extracting the critical exponent $\nu$ from a systematic finite size scaling analysis of the Binder cumulant of the 
    imaginary part of the \PL. We are able to explain a previously observed finite size effect afflicting the 
    scaling of the Binder cumulant in the regime of three-phase coexistence.
    Compared to $\NTau=4$ lattices, the tricritical masses are significantly shifted.
  Exploratory results on $\NTau=8$ as well as comparison with staggered simulations suggest that 
  much finer lattices are needed before a continuum extrapolation becomes feasible.
\end{abstract}

\pacs{12.38.Gc, 05.70.Fh, 11.15.Ha}
\keywords{QCD phase diagram, Binder cumulant}
\maketitle


\section{Introduction}\label{ch:introduction}
One of the most challenging aspects of modern particle physics is to map out the phase diagram of Quantum Chromodynamics (QCD)  as a function of temperature $T$ and baryon chemical potential $\mu_B$.
Due to the non-perturbative nature of the strong interactions on hadronic energy scales, a first principles approach such as Lattice QCD (LQCD) is mandatory.

At zero baryon chemical potential, standard Monte Carlo simulations can be applied.
In order to understand the interplay between confinement and chiral symmetry breaking and their influence on the thermal transition, it is interesting to study the QCD phase diagram varying the quark masses between the chiral ($m\to 0$) and quenched ($m\to\infty$) limits.
For $\Nf=2,3$ degenerate quark flavours, regions of first-order chiral and deconfinement transitions are seen on coarse $\NTau=4,6$ lattices with standard actions for light and heavy quark masses, respectively, whereas intermediate mass regions including the physical point show crossover behaviour.
For improved actions, the chiral first order region is significantly smaller, but presently no continuum extrapolation of any of these features is available (see Ref.~\onlinecite{Meyer:2015} and references therein for a recent overview).

At finite $\mu_B$, the sign problem prevents importance sampling techniques and alternative strategies must be used.
One possibility is to introduce a purely imaginary quark chemical potential $\mu\equiv\mu_B/3=\imath\MuI$ ($\MuI\in\mathbb{R}$), for which no sign problem is present.
The phase structure at imaginary chemical potential constrains the situation at real $\mu_B$ by analytic continuation.

In the last decade, a first understanding of the QCD phase diagram at imaginary chemical potential has been developed as summarized in \sectionname~\ref{sec:QCDpd}.
It is so far based on investigations on coarse lattices ($\NTau=4$, $a\sim 0.3$ fm) with staggered 
fermions~\cite{deForcrand:2010he,D'Elia:2009qz,Bonati:2010gi}
and standard \cite{Philipsen:2014rpa} or improved \cite{Alexandru:2013uaa} Wilson fermions only.
In the present work, we repeat the study made in \referencename~\onlinecite{Philipsen:2014rpa} on a finer lattice ($\NTau=6$, $a\sim 0.2$ fm). 
Unfortunately, we find that several further and more costly simulations are required before 
any continuum extrapolation can be attempted.

After a brief description of the QCD phase diagram in \sectionname~\ref{sec:QCDpd}, we illustrate our simulation
setup in \sectionname~\ref{sec:numericSetup}. \sectionname~\ref{sec:bump} is dedicated to a study of the qualitative behaviour of the Binder cumulant, which explains some puzzling finite size effects observed in 
earlier studies.
The results of our investigation are presented and discussed in \sectionname~\ref{sec:results}.

\section{QCD phase diagram at imaginary chemical potential}\label{sec:QCDpd}
The QCD phase diagram for purely imaginary values of the chemical potential $\mu=\imath\MuI$ has a rich structure that depends on the temperature $T$, chemical potential $\MuI$ as well as  on the number of flavours and the 
values of the quark masses.

\begin{figure}[t]
  \centering
  \includegraphics[scale=0.5]{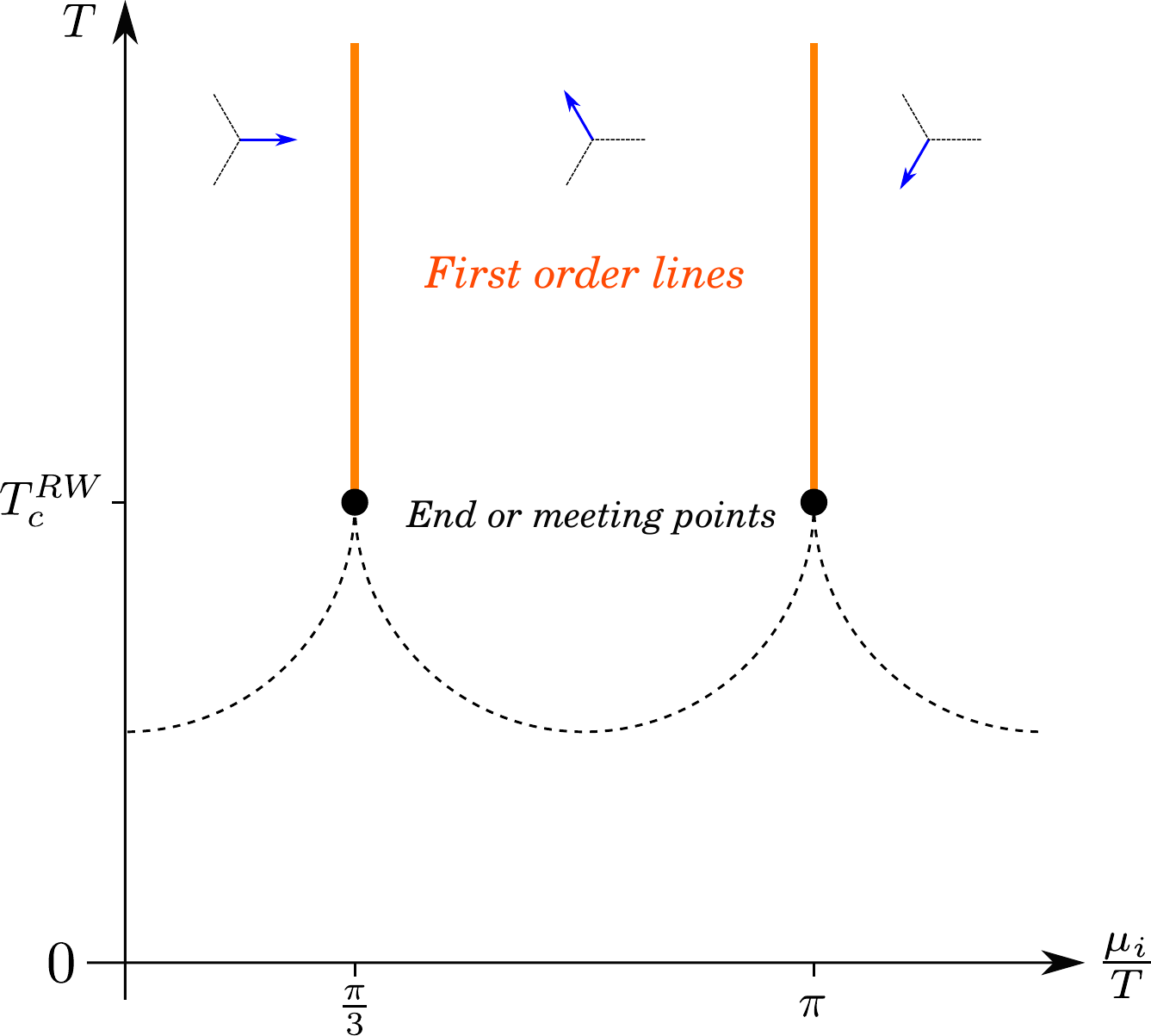}
  \caption{QCD phase diagram in the $T-\MuI$ plane. The dashed line depicts the chiral/deconfinement
           transition whose nature depends on the quark masses. The orange lines represent the Roberge-Weiss
           (RW) transitions. The black dots, where the first-order lines terminate, can be first-order
           triple points, tricritical points or second-order endpoints.}
  \label{fig:RW_T_mu_plane}
\end{figure}

\begin{figure}[t]
  \centering
  \includegraphics[scale=0.5]{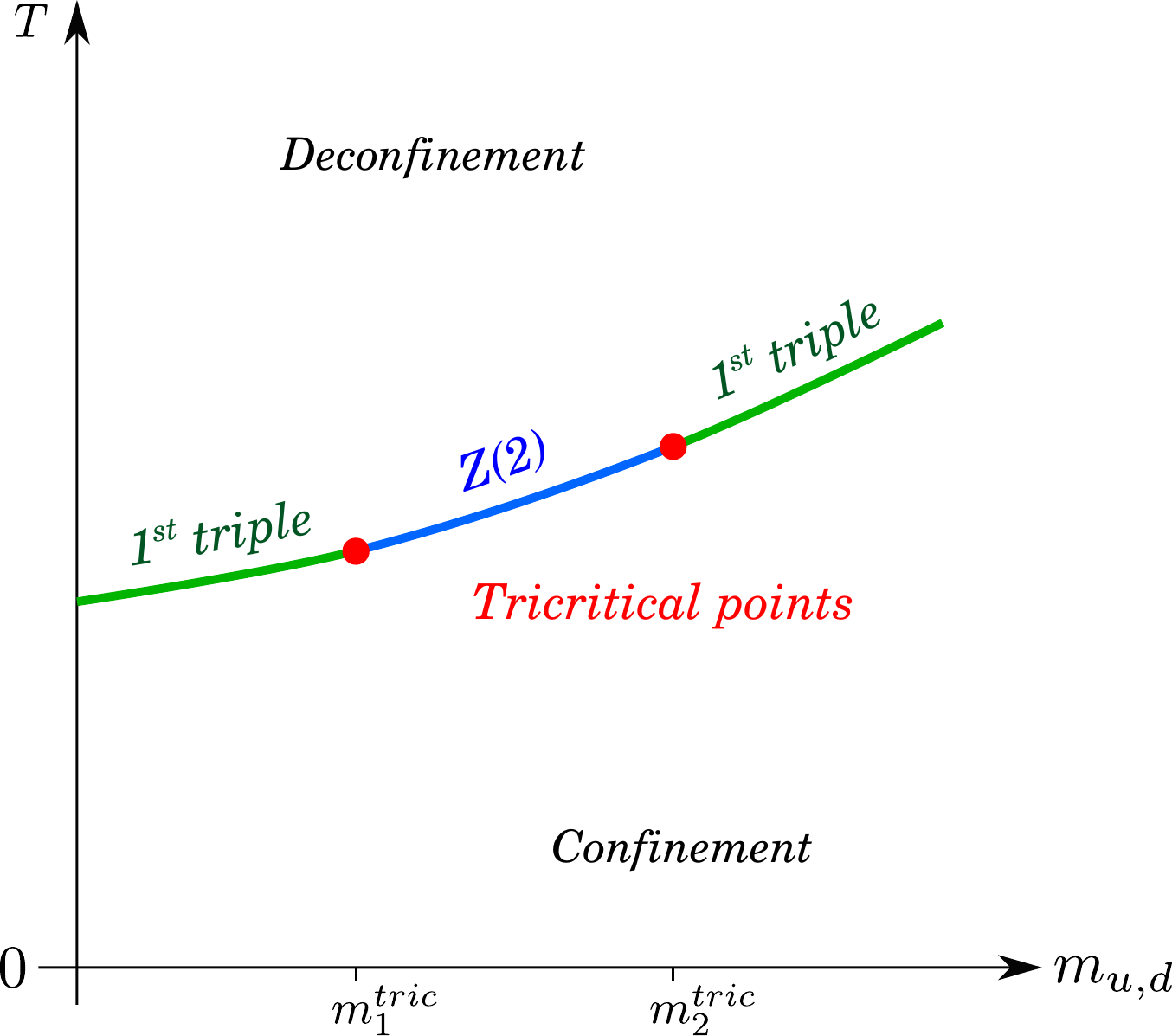}
  \caption{QCD phase diagram in the $T-m_{u,d}$ plane for a fixed critical value of the imaginary 
           critical potential $\MuI=\MuI^c$.}
  \label{fig:RW_T_mass_plane}
\end{figure}

The QCD partition function is symmetric by reflection in $\mu$ and it is periodic in $\MuI/T$ with period $2\pi/\Nc$~\cite{Roberge:1986mm}.
These two properties imply the phase structure depicted qualitatively in \figurename~\ref{fig:RW_T_mu_plane} (from now on we fix $\Nc=3$). 
In particular, varying the imaginary chemical potential, phase transitions between different $Z(3)$ sectors are crossed at fixed values $\MuI^c/T=(2k+1)\pi/3$ with $k\in\mathbb{Z}$ (the so called \RW{} transitions).
Such transitions are smooth crossovers for low $T$ and true first-order phase transitions for high $T$~\cite{Roberge:1986mm}.
Any physical observable is invariant under a 
change of the $Z(3)$ centre sector (i.e. shifting $\MuI/T$ by its period), which can be distinguished by the phase of the \PL{} $L$.
For any spatial lattice site $\mathbf{n}$, 
\begin{equation}\label{eq:PolyakovLoop}
    L(\mathbf{n})=\frac{1}{3}\Tr_C \Biggl[\prod_{n_0=0}^{N_\tau-1}\Link_0(n_0,\mathbf{n})\Biggr]
                 \equiv \lvert L(\mathbf{n})\rvert e^{-\imath\varphi} \;,
\end{equation}
where, as different sectors are explored, the phase $\varphi$ takes the values $\langle\varphi\rangle=2n\pi/3$ with $n\in\{0,1,2\}$. 
The dashed line in \figurename~\ref{fig:RW_T_mu_plane} represents the analytic continuation of the 
chiral/deconfinement transition which is crossed varying the temperature.
Its type depends on the values of the quark masses.
Consequently, also the nature of the meeting points of the dashed line and the first-order RW lines is mass--dependent.
Recent studies~\cite{deForcrand:2010he,D'Elia:2009qz,Bonati:2010gi} show that, for $\Nf=2$ and $\Nf=3$ on coarse lattices, these points are first-order triple points for small and large masses, while they are second-order endpoints for intermediate masses. Therefore, there are two tricritical points separating the two regimes. This has been schematically drawn in \figurename~\ref{fig:RW_T_mass_plane}.

\begin{figure*}[t]
  \centering
  \includegraphics[scale=0.7]{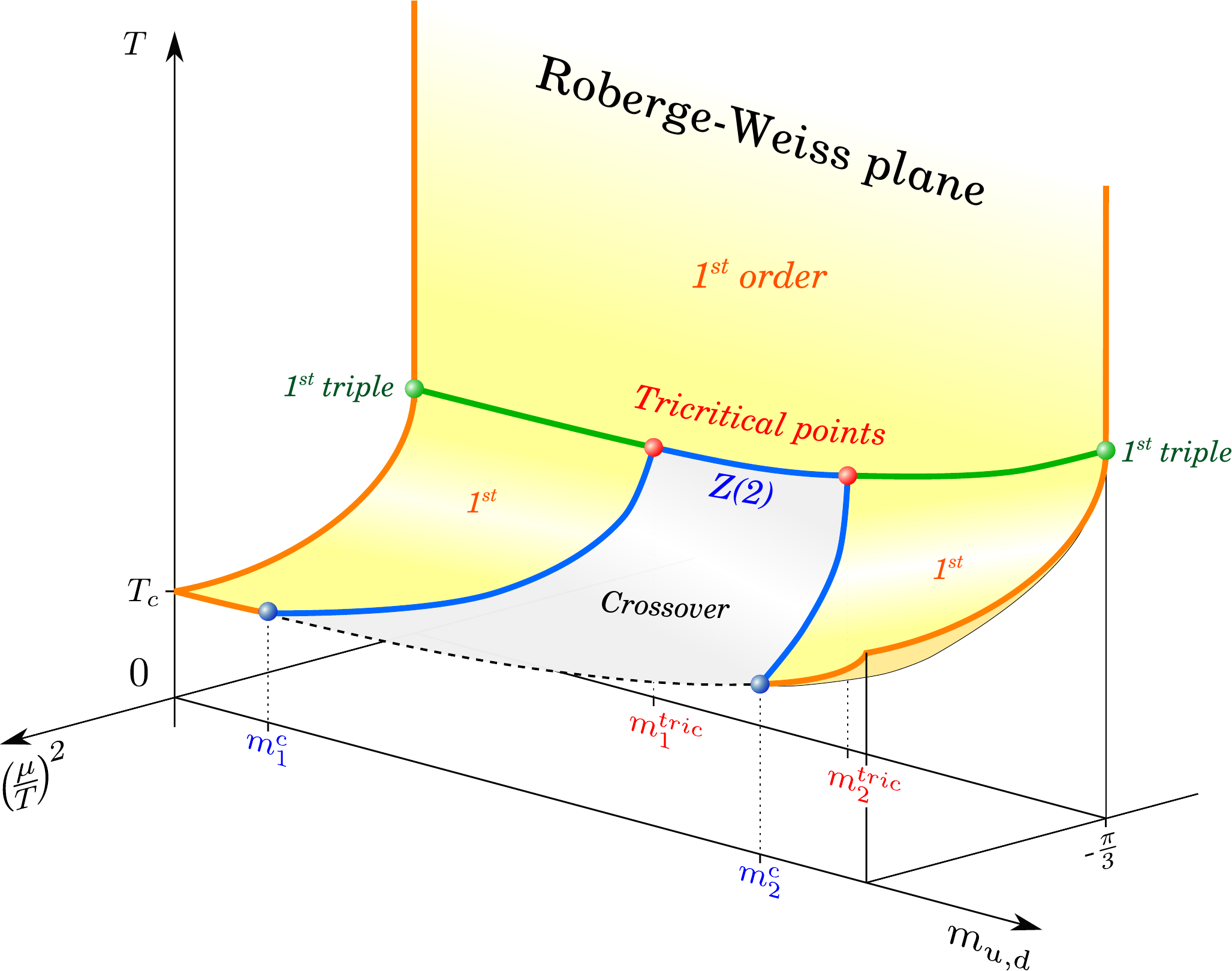}
  \caption{$\Nf=2$ QCD phase diagram in the $T-\mu-m_{u,d}$ space for
           $\displaystyle -\frac{\pi}{3}\le\Bigl(\frac{\mu}{T}\Bigr)^2\le 0$.}
  \label{fig:RW_T_mu_mass_diagram}
\end{figure*}

\figurename~\ref{fig:RW_T_mu_mass_diagram}
combines \figurename~\ref{fig:RW_T_mu_plane} and \figurename~\ref{fig:RW_T_mass_plane} into a 3D picture. 
On coarse lattices, the first-order chiral transition region extends through $\mu=0$, producing a
critical point $m_1^c$ in the $T-m_{u,d}$ plane \cite{Bonati:2013tqa,Philipsen:2015eya}.
Slicing \figurename~\ref{fig:RW_T_mu_mass_diagram} at $m_{u,d}=\text{const.}$ allows to understand how the nature of the dashed line of \figurename~\ref{fig:RW_T_mu_plane} changes.
\figurename~\ref{fig:RW_T_mu_mass_diagram} has been drawn for $0<\MuI<\pi/3$, the situation at any other value of $\MuI$ can be deduced using the symmetries of the partition function.
Note that the position of the (tri)critical points and thus also the shape of the $Z(2)$ lines changes as the continuum limit is approached.
Reducing the lattice spacing, the low mass first-order region shrinks \cite{deForcrand:2007rq}, 
while the high mass one enlarges~\cite{Fromm:2011qi}.
Similarly, the tricritical masses measured in physical units on $\NTau=4$ lattices have rather different
values in different fermion discretizations~\cite{Bonati:2010gi,Philipsen:2015eya}.
The present work is a first step towards understanding the cut-off effects in the Wilson formulation.

\section{Simulation setup}\label{sec:numericSetup}
After performing the integration over the fermionic fields, the \Qcd{} grand-canonical partition function with $\Nf=2$ mass-degenerate quarks in presence of an imaginary chemical potential $\MuI$ reads
\[
    \PartFunc(T,\MuI)=\int \mathcal{D}\Link e^{-\SGluon[\Link]}\;\bigl(\det D[\Link,\MuI]\bigr)^2 \;,
\]
where $\SGluon$ is the gauge part of the action and $D$ is the fermion matrix. For our study we used the standard Wilson gauge action,
\[
    \SGluon[\Link]= \beta \sum_{P}\Bigl\{ 1 - \Re \bigl[\Tr_C P\bigr]\Bigr\} \;,
\]
and the standard Wilson discretization of dynamical fermions, with the fermion matrix 
\[
    D_{i,j} = \delta_{i,j}
            - \kappa \!\sum_{\rho=\pm 0}^{\pm 3}
                   e^{\imath a \MuI \cdot \delta_{|\rho|,0}\cdot\sign(\rho)}
                   \Bigl[(1-\gamma_\rho)\,\Link_\rho(i)\,\delta_{i+\hat{\rho},j} \Bigr] \;.
\]
In the last two equations, $\beta$ is the lattice coupling (related to the bare coupling $g$ via $\beta=6/g^2$), $P$ indicates the plaquette, $i$ and $j$ refer to lattice sites, $\hat{\rho}$ is a unit vector on the lattice and $a$ is the lattice spacing. Moreover $\gamma_{-\rho}\equiv-\gamma_\rho$ and $\Link_{-\rho}(i)\equiv\Link^\dag_\rho(i-\vec{\rho})$. The bare quark mass $m_{u,d}\equiv m$ is contained in the hopping parameter $\kappa$ via
\[
    \kappa=\frac{1}{2(a\,m+4)}\;.
\]

The shifted phase $\phi=\varphi-\MuI/T$ of the \PL{} is an order parameter to distinguish between the low $T$ disordered phase and the high $T$ ordered phase with two-state coexistence \citep{deForcrand:2010he}.
For the particular, critical values $\MuI/T=\pi\pm 2\pi k,\;k\in\mathbb{Z}$, also the imaginary part of the \PL{} behaves as an order parameter. This is the reason why we fixed $\MuI/T=\pi$ in all our simulations.
Since the temperature on the lattice is given by
\[
    T=\frac{1}{a(\beta)\,N_\tau}\;,
\]
we have $a\MuI=\pi/6$ for $N_\tau=6$.

In order to identify the nature of the \RW{} end- or meeting point, 
we use the Binder cumulant \cite{Binder:1981sa} defined as
\[
    \Binder(X,\alpha_1,\dots,\alpha_n)
         \equiv\frac{\bigl\langle(X-\langle X\rangle)^4\bigr\rangle}
                    {\bigl\langle(X-\langle X\rangle)^2\bigr\rangle^2 \vphantom{\Bigl[}} \;,
\]
where $X$ is a general observable and $\alpha_1,\dots,\alpha_n$ is a set of parameters on which $\Binder$ depends. Critical parameter values $\alpha_i^c$ are defined by the vanishing of the third moment of the fluctuations. 
In the thermodynamic limit $V\to\infty$, i.e. when non-analytic phase transitions can exist, the Binder cumulant  evaluated at critical couplings then takes different values depending on the nature of the phase transition (see \tablename~\ref{tab:BinderValues}).

In our study we choose $X=\PolyIm$ (in the following $\Poly$ stands for the spatially averaged $L(\mathbf{n})$ of \Eq{eq:PolyakovLoop}) and $\{\alpha_i\}=\{\beta,\kappa,\MuI\}$.
Since we work at the critical value $\MuI=\pi\,T$, then, at any value of the temperature, $\langle(X-\langle X\rangle)^3\rangle\approx0$ and we expect the Binder cumulant to be close to 3 (crossover) for low $T$ and close to 1 (first order) for high $T$.
Even though \Binder{} is a non-analytic step function for $V\to\infty$, at finite volume it gets smoothed out and its slope increases with the volume.
Around the critical coupling $\beta_c$, the Binder cumulant is expected to show a well-defined finite size scaling behaviour.
It is then a function of $x\equiv(\beta-\beta_c)N_\sigma^{1/\nu}$ only and can be Taylor-expanded as
\begin{equation}\label{eq:BinderFSS}
    \Binder(\beta, N_{\sigma}) = \Binder(\beta_c,\infty) + a_1\,x + a_2\,x^2 + \mathcal{O}(x^3) \;.
\end{equation}
Close to the thermodynamic limit, the intersection of different volumes gives $\beta_c$ 
and the critical exponent $\nu$ takes its universal value depending on the type of transition.
In \tablename~\ref{tab:BinderValues} the values of the critical exponents relevant for our work have been summarized~\cite{Pelissetto:2000ek}.

Another important quantity is the order parameter susceptibility, defined as
\[
    \chi(X)\equiv N_\sigma^3 \bigl\langle(X-\langle X\rangle)^2\bigr\rangle   \;.
\]
Also this quantity is expected to scale around $\beta_c$ according to
\begin{equation}\label{eq:SuscScaling}
    \chi=N_\sigma^{\gamma/\nu} f(t\,N_\sigma^{1/\nu})   \;,
\end{equation}
where $t\equiv(T-T_c)/T_c$ is the reduced temperature and $f$ a universal scaling function.
This means that, once the critical exponents $\gamma$ and $\nu$ are fixed to the correct values, $\chi/N_\sigma^{\gamma/\nu}$ measured on different lattice sizes should collapse when plotted against $t\,N_\sigma^{1/\nu}$. We also performed occasional cross-checks of the susceptibility for $X=\chiralcond$ leading to fully consistent results.

Our strategy to locate the two tricritical values of $\kappa$ is completely analogous to that 
used in \referencename~\onlinecite{Philipsen:2014rpa}.
For each simulated value of $\kappa$, we measured the Binder cumulant in the critical region and extracted the values of $\Binder(\beta_c,\infty)$, $a_1$, $\beta_c$ and $\nu$ fitting our data according to \Eq{eq:BinderFSS}, considering the linear term only.
The changes in $\nu$ as $\kappa$ is varied allow to locate the tricritical points.

We studied 9 values of the bare quark mass between $\kappa=0.1$ and $\kappa=0.165$.
For each value of $\kappa$, we simulated at the fixed temporal lattice extent $N_\tau=6$ that implies the value $a\MuI=\pi/6$ for the imaginary chemical potential.
Three or four different spatial lattice sizes per $\kappa$ have been used, always with $N_\sigma\geq16$ (except for $\kappa=0.1625$ where also $N_\sigma=12$ was used).
This gives a minimal aspect ratio of almost 3.
For every lattice size, 6 up to 30 values of $\beta$ around the critical value have been simulated.
Between 40k--500k standard HMC~\cite{Duane:1987de} trajectories of unit length per $\beta$ have been collected after at least 5k trajectories of thermalization.
The observables of interest (i.e. plaquette, $\PolyRe$ and $\PolyIm$) were measured for every trajectory after the thermalization.
In each run the acceptance rate was tuned to $\sim$75\%.
For $\kappa\geq 0.16$, i.e. for the smallest masses, the Hasenbusch trick~\cite{Hasenbusch:2001ne} in the integration of the Molecular Dynamics equations has been used to reduce the integrator instability, which is triggered by isolated small modes of the fermion kernel~\cite{Joo:2000dh}.
Because of the particularly delicate fitting procedure required to extract the critical exponent $\nu$ from \Eq{eq:BinderFSS}, we almost always produced 4 different Markov chains for each value of the coupling in order to better understand if the collected statistics was enough.
Ferrenberg-Swendsen reweighting~\cite{Ferrenberg:1989ui} was used to smoothly interpolate between
$\beta$-points (see \appendixname~\ref{app:nu} for more information about the method used to extract $\nu$, \appendixname~\ref{app:sim} for the simulations details).

For scale-setting purposes, $\Temp=0$ simulations at or close to certain critical parameters have been performed.
$\Order(400)$ independent configurations on $16^3\times32$ lattices have been produced.
The scale itself is then set by the Wilson flow parameter $w_0$ using the publicly available code described in \referencename~\onlinecite{Borsanyi:2012zs}.
This method is very efficient and fast.
In addition, the pion mass \mpi\ was determined using these configurations.
See Table \ref{tab:scaleSetting} for more details.

All our numerical simulations have been performed using the \emph{publicly available}~\cite{CL2QCD} \Ocl~\cite{opencl} based code \clqcd~\cite{Bach:2012iw,Philipsen:2014mra}, which is optimized to run efficiently on GPUs. In particular, the \Loewe~\cite{Bach2011a} at Goethe-University Frankfurt and the \Lcsc~\cite{L-CSC} at GSI in Darmstadt have been used.

\begin{table}[t]
  \centering
  \[
  \begin{array}{*{5}{c}}
    \toprule[0.3mm]
    & \text{Crossover} & 1^{st} \text{ triple} & \text{Tricritical} & 3D \text{ Ising} \\
    \midrule[0.1mm]
    \Binder      & 3 & 1.5 & 2   & 1.604     \\
    \nu          & - & 1/3 & 1/2 & 0.6301(4) \\
    \gamma       & - & 1   & 1   & 1.2372(5) \\
    \bottomrule[0.3mm]
  \end{array}
  \]
  \caption{Critical values of $\nu$, $\gamma$ and $\Binder\equiv\Binder(X,\alpha_c)$
           for some universality classes~\cite{Pelissetto:2000ek}.}
  \label{tab:BinderValues}
\end{table}

\section{The Binder cumulant Bump}\label{sec:bump}

\begin{figure*}
    \centering
    \subfigure[Binder cumulant reweighted data for $\kappa=0.165$.]
    {\label{fig:k1650bump}\includegraphics[width=.45\textwidth]{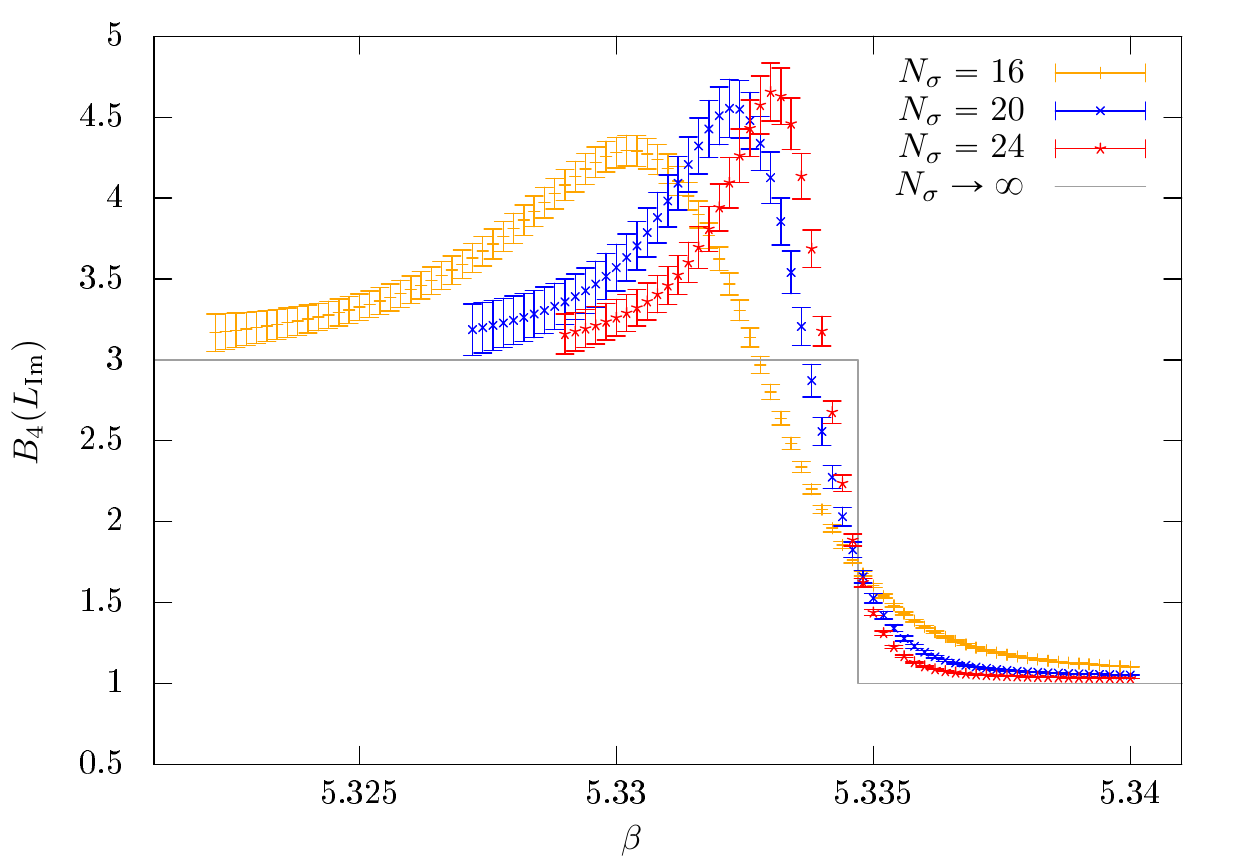}} \qquad
    \subfigure[Binder cumulant as function of $\beta$ in our model.]
    {\label{fig:binderBump2D}\includegraphics[width=.45\textwidth]{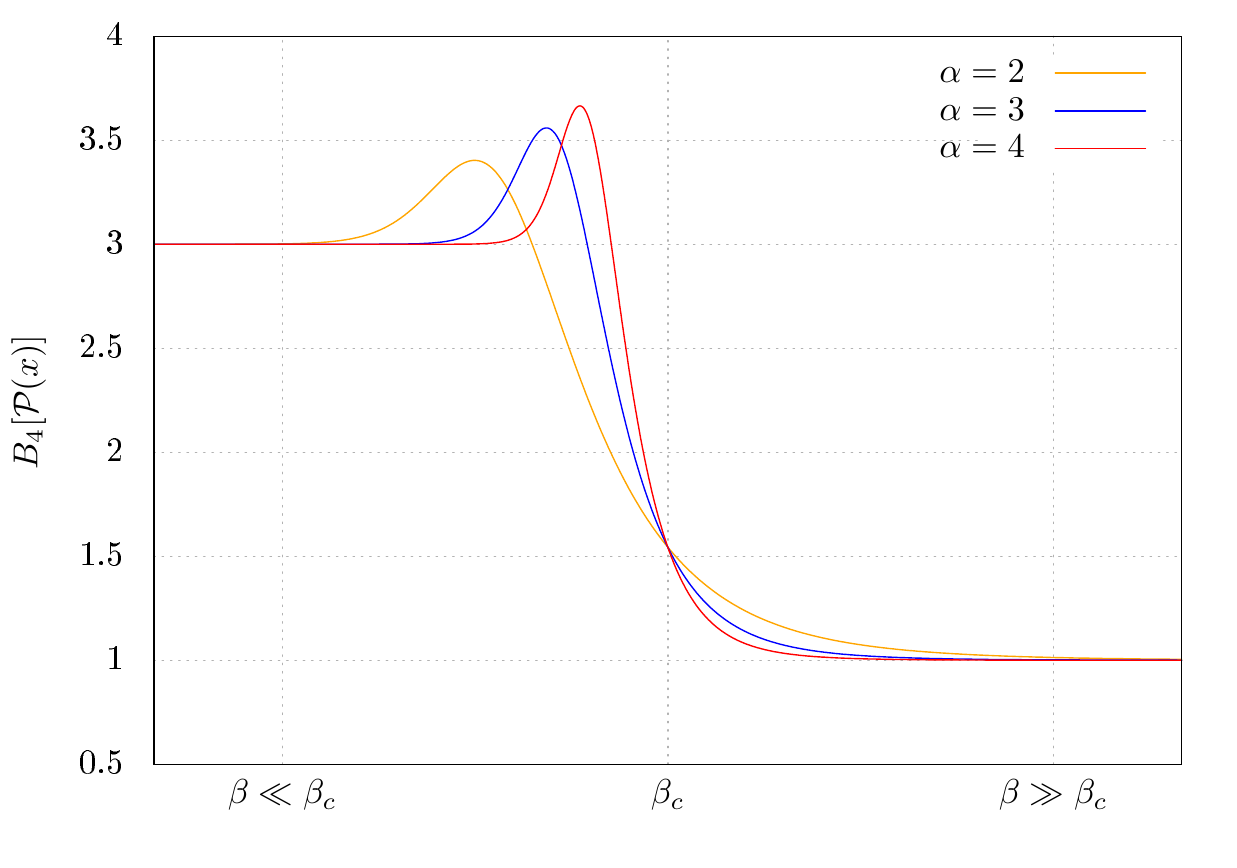}}
    \caption{Comparison between the measured Binder cumulant of the imaginary part of the \PL{}
             and its analytic form in our model. Three different lattice spatial extents and three different
             values of the parameter $\alpha$ have been used. The gray line in the left plot is the expected
             behaviour of $\Binder(\beta)$ in the thermodynamic limit.}
    \label{fig:binderBumps}
\end{figure*}

As explained in \sectionname~\ref{sec:numericSetup}, the Binder cumulant is expected to change from 3 at low $T$ to 1 at high $T$.
It is also known that $\Binder(\beta)=2\;\Theta(\beta_c-\beta)+1$ in the thermodynamic limit, where $\Theta$ is the Heaviside step function.
On finite volumes the discontinuity is smoothed out and the Binder cumulant could naively be expected to be a monotonic function of $\beta$.
However, it turns out that $\Binder$ takes values higher than 3 at $\beta\lesssim\beta_c$ for small and large values of $\kappa$, i.e. in the first-order regions.
In \figurename~\ref{fig:k1650bump} the data for $\kappa=0.165$ are shown,
with a ``bump'' rising to values significantly larger than 3 on the crossover side of the transition. 
Note how the bump gets higher and narrower on larger volumes.
Moreover, the $\beta$-region where $\Binder$ changes from 3 to 1 shrinks as $\NSigma$ is increased, as expected for a first-order transition.
The occurrence of the bump has been reported also in other studies~\cite{Alexandru:2013uaa}. 
This distorts the finite size analysis compared to the naive expectations, and in particular leads to significantly 
higher values of the Binder cumulant at the intersection than expected in the thermodynamic limit 
\cite{ deForcrand:2010he,Alexandru:2013uaa,Philipsen:2014rpa}. Thus, the effect needs to be understood 
if one aims at results in the thermodynamic limit.

\begin{figure*}
  \centering
  \subfigure[Symmetrized histograms of $\PolyIm$ for $\kappa=0.165$ and $\NSigma=24$ at various values of $\beta$ around $\beta_c$.]
  {\label{fig:distributionData}\includegraphics[width=.45\textwidth]{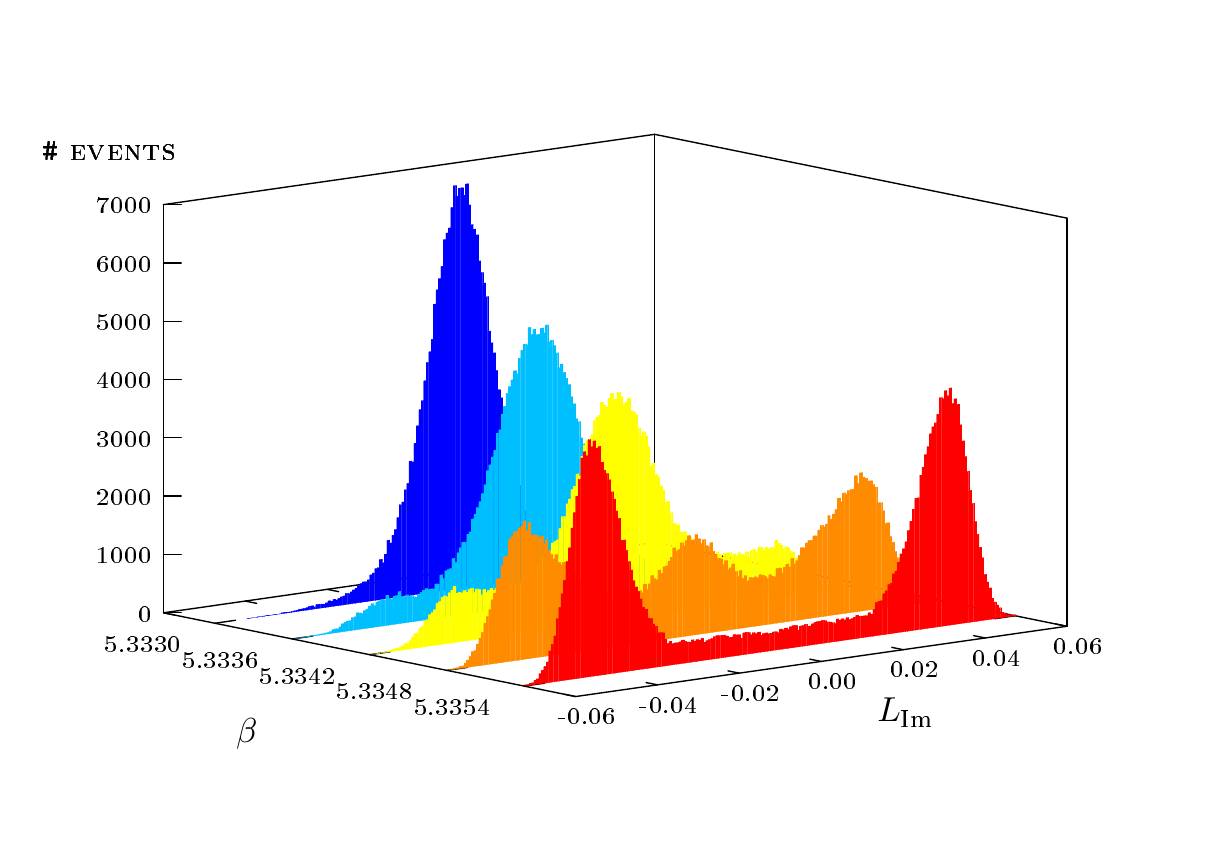}} \qquad
  \subfigure[Modelled probability distribution of \Eq{eq:distribution}, plotted as function of $d$ and $x$ for fixed $\sigma=0.1$ and $\alpha=1$.]
  {\label{fig:distribution}\includegraphics[width=.45\textwidth]{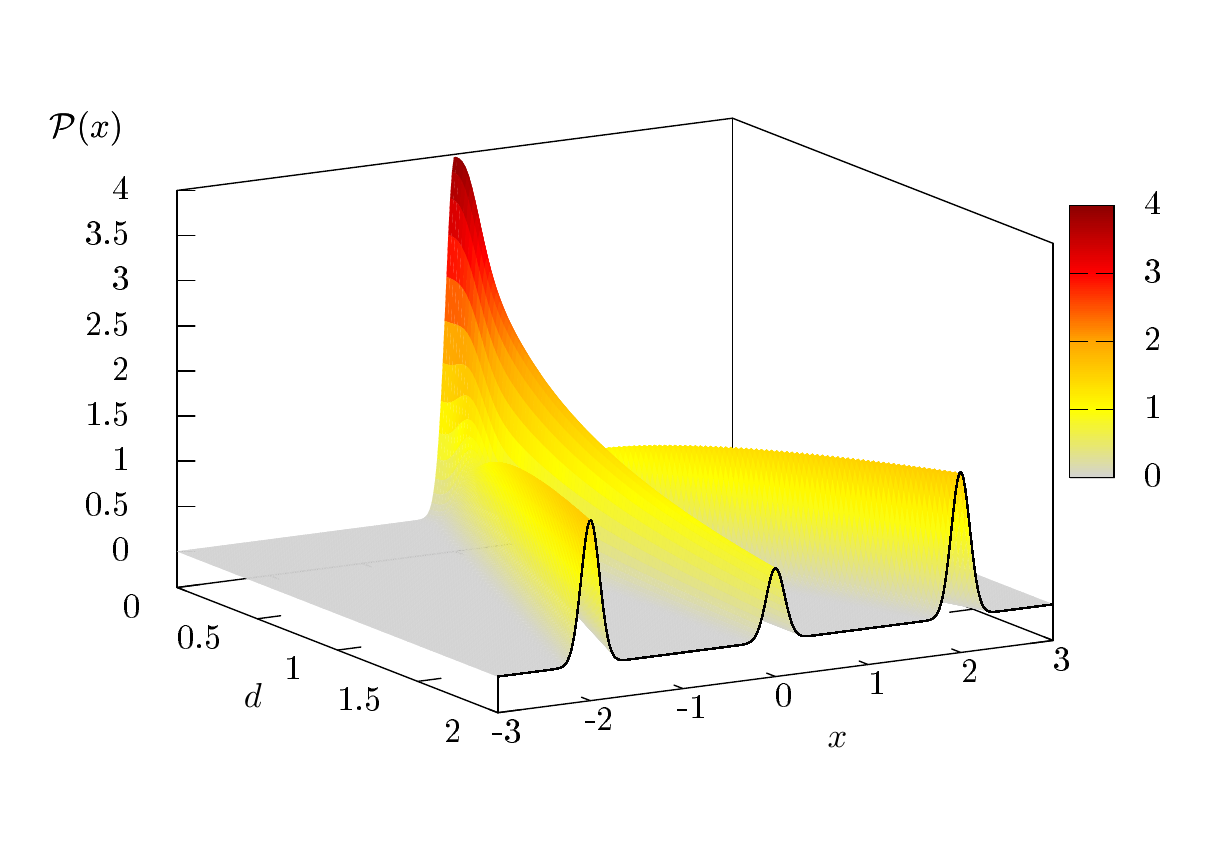}}
  \caption{Comparison between histograms of the imaginary part of the \PL{} and the corresponding probability distribution in the model.
           Note how $d$ is related to $\beta$.
           For low(high) temperatures, only one(two) Gaussian distribution(s) is(are) present.
           Moreover, in (\subref{fig:distributionData}) at $\beta=5.3348$ (i.e. slightly after the transition), a clear three-peak structure is visible,
           as expected for a triple point.
           All these features are captured in the model.}
\end{figure*}

The described behaviour can be explained by modelling the distributions at work in a situation with three phases.
Let us consider the distribution of the imaginary part of the \PL{} on a finite volume for sufficiently high statistics: it is a normal distribution for $\beta\ll\beta_c$ (crossover) and it is the sum of two normal distributions with mean values $\pm\lvert\PolyIm\rvert$ for $\beta\gg\beta_c$ (first order).
This is clearly visible in \figurename~\ref{fig:distributionData}, where histograms of $\PolyIm$ are depicted.
Around the transition, the $\PolyIm$ distribution can be thought of as the sum of three Gaussian distributions, whose weights depend on the temperature. We thus consider
\begin{equation}\label{eq:distribution}
    \mathcal{P}(x)\equiv w_o\;\mathcal{N}(-d,\sigma) + w_i\;\mathcal{N}(0,\sigma)+w_o\;\mathcal{N}(d,\sigma)\;,
\end{equation}
where
\[
    \mathcal{N}(\mu,\sigma)\equiv\frac{1}{\sigma\sqrt{2\pi}}\;e^{-\frac{(x-\mu)^2}{2\sigma^2}}
\]
is a Gaussian distribution with mean $\mu$ and variance $\sigma^2$, $d$ is a positive real number, while $w_o$ and $w_i$ are the weights of the outer and inner distributions, respectively.
Of course, $2 w_o+w_i=1$.
Here, for simplicity, we assumed the three distributions to have the same variance.
The symmetry of the outer distributions with respect to zero and the fact that their weight is the same are, instead, implied by the symmetries of the physical system.
It is clear that $d$ has to be a function of $\beta$ as well as $w_o$ and $w_i$.
In particular, we have $w_o\approx0$ and $d\approx0$ for $\beta\ll\beta_c$,
while $w_i\approx0$ and $d\gg\sigma$ , i.e. the outer Gaussian distributions are well separated,  for $\beta\gg\beta_c$.
With an analytic expression for the distribution, the value of the Binder cumulant for an even function can be explicitly calculated through 
\[
    \Binder\bigl[\mathcal{P}(x)\bigr]=\frac{\int_{-\infty}^{+\infty} x^4\;\mathcal{P}(x)\;dx}
                                                 {\Bigl[\int_{-\infty}^{+\infty} x^2\;\mathcal{P}(x)\;dx\Bigr]^2}\;
\]
and we will have indeed
\begin{subequations}\label{eq:BinderLowHighT}
    \begin{equation}
        \Binder\bigl[\mathcal{P}_{\beta\ll\beta_c}(x)\bigr]=3   
    \end{equation}
while
    \begin{equation}
        \Binder\bigl[\mathcal{P}_{\beta\gg\beta_c}(x)\bigr]=3-\frac{2 d^4}{(d^2+\sigma^2)^2}\approx1\;. 
    \end{equation}
\end{subequations}
Before trying to further connect our parameters $d$, $w_o$ and $w_i$ to $\beta$, let us just study how the Binder cumulant of our distribution changes as they are varied.
At the end of the section, we will comment further on how the quantities in our simple model are related to the physical ones.

It is possible to think of the two cases in \Eqs{eq:BinderLowHighT} as the two limits $d\to0$ and $d\to\infty$, on condition that the weights of the distributions modify accordingly.
One way to realize this is to assume that both $w_o$ and $w_i$ are functions of $d$, satisfying
 the following conditions:
\begin{align*}
    \lim_{d\to\,0}w_i(d)=1 \quad&\text{and}\quad \lim_{d\to\,0}w_o(d)=0\;;\\
    \lim_{d\to\,\infty}w_i(d)=0 \quad&\text{and}\quad \lim_{d\to\,\infty}w_o(d)=\frac{1}{2}\;.
\end{align*}
Now, in order to properly model the weights to reproduce the bump of \figurename~\ref{fig:k1650bump}, we first
have to  understand how a Binder cumulant larger than~3 can arise.
Leaving the weights of the three normal distributions completely general, it can be shown that
\[
    \Binder\bigl[\mathcal{P}(x)\bigr]=3+\frac{2\,w_o\,d^4\:(w_i-4\,w_o)}{(2\,w_o\,d^2+\sigma^2)^2}\;.
\]
Hence, when the weight of the central distribution is more than 4 times larger than the weight of the outer distributions, the Binder cumulant takes values larger than 3.
It is then sufficient to choose the functions $w_o(d)$ and $w_i(d)$ to respect the limits above and in a way such that
\begin{equation}\label{eq:bumpCondition}
    w_i(d)>4\:w_o(d)
\end{equation}
for some values of $d$. A simple choice to respect the required asymptotic behaviour is
\begin{subequations}\label{eq:weights}
    \begin{align}
        w_i(d)&=\frac{\frac{1}{\alpha d+1}}{\frac{1}{\alpha d+1}+2\Bigl(1-\frac{1}{\frac{d}{\alpha}+1}\Bigr)}
               =\frac{\alpha+d}{\alpha+3d+2\alpha d^2}\;,\\[1ex]
        w_o(d)&=\frac{1-\frac{1}{\frac{d}{\alpha}+1}}{\frac{1}{\alpha d+1}+2\Bigl(1-\frac{1}{\frac{d}{\alpha}+1}\Bigr)}
               =\frac{d\:(1+\alpha d)}{\alpha+3d+2\alpha d^2}\;,
    \end{align}
\end{subequations}
where $\alpha>0$ is a parameter to calibrate how fast the weights $w_i(d)$ and $w_o(d)$ change from 1 to 0 and from 0 to 1/2, respectively.
More precisely, the larger $\alpha$ the quicker the inner(outer) Gaussian distribution(s) disappears(appear).
In \figurename~\ref{fig:distribution} it is shown how the distribution $\mathcal{P}(x)$ changes increasing the parameter $d$ for $\sigma=0.1$ and $\alpha=1$.
One clearly sees that for small $d$ there is almost only the inner Gaussian.
For higher $d$, the middle normal distribution gradually disappears. Thus $d$ plays the role of temperature
or $\beta$, and $\alpha$ that of the volume.

\begin{figure}[t]
  \centering
  \includegraphics[scale=0.7]{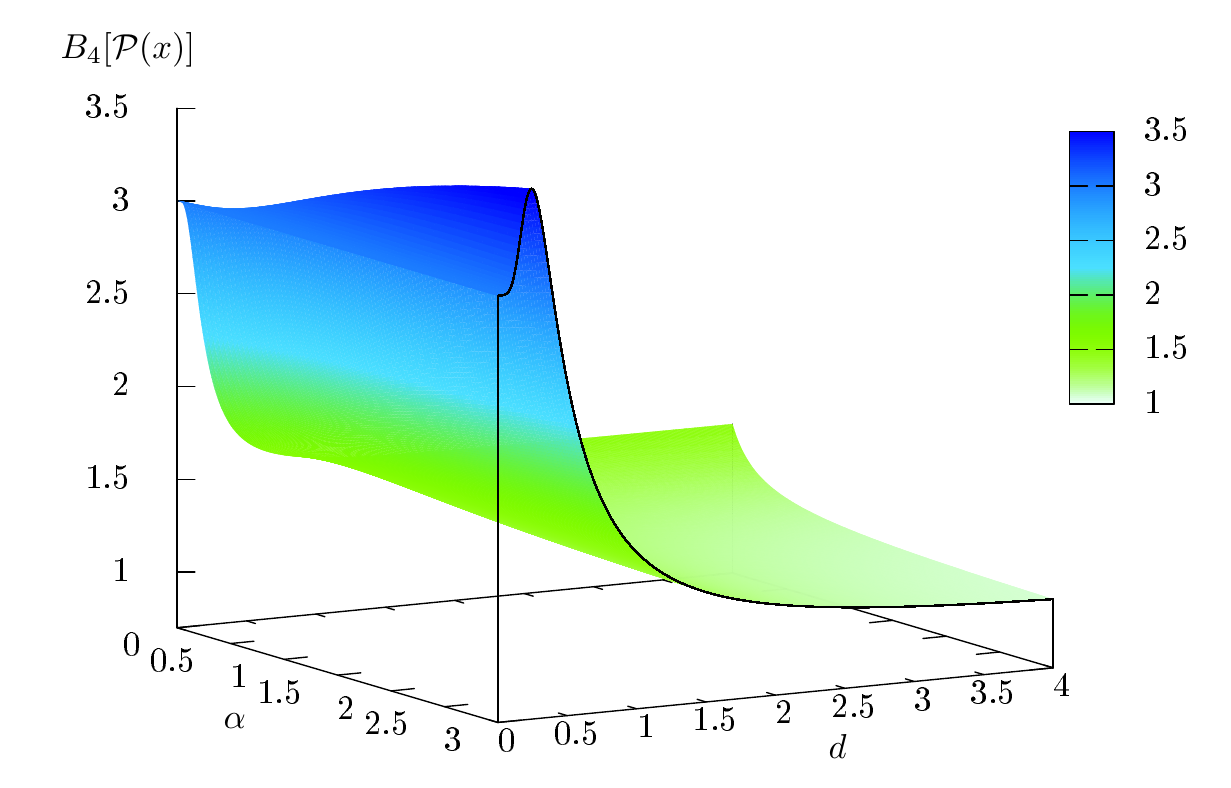}
  \caption{Binder cumulant of the distribution in \Eq{eq:distribution}
           for $\sigma=0.1$ with the weights of \Eqs{eq:weights}.}
  \label{fig:binderBump3D}
\end{figure}

The region where the Binder cumulant is larger than 3 can be found by inserting \Eqs{eq:weights} in \Eq{eq:bumpCondition}. Then, it follows that
\begin{equation}\label{eq:BinderBigger3}
    \Binder>3 \quad\Leftrightarrow\quad0<d<\frac{-3+\sqrt{9+16\alpha^2}}{8\alpha}\:;
\end{equation}
actually, using the chosen weights in \Eq{eq:distribution}, we get
\[
    \Binder\bigl[\mathcal{P}(x)\bigr]=3-\frac{2\:d^5\:(1+\alpha d)\:(4\alpha d^2+3d-\alpha)}
                                             {[2\:d^3\:(1+\alpha d)+\sigma^2\:(\alpha+3d+2\alpha d^2)]^2}\;,
\]
which confirms what is expected in \Eq{eq:BinderBigger3}.
In \figurename~\ref{fig:binderBump3D} the Binder cumulant of the distribution $\mathcal{P}(x)$ is plotted as function of $\alpha$ and $d$, keeping the standard deviation $\sigma$ fixed.
This picture qualitatively describes our data, as can be seen comparing it to \figurename~\ref{fig:k1650bump}. In particular, the height/width of the bump increases/shrinks as the parameter $\alpha$ is increased.

Lastly, we give some remarks about the connection between $d$ and the temperature.
As already observed, it has certainly to be that $d=d(\beta)$.
This function should reproduce the fact that the Binder cumulant stays on the value 3 for $\beta\ll\beta_c$, it should let the bump occur for $\beta\lesssim\beta_c$ and it should make the Binder take the correct value for $\beta=\beta_c$.
Since we know that $\Binder$ is 3 for $d\to 0$, then the first aspect can be reproduced choosing a function of $\beta$ that is almost zero for $\beta\ll\beta_c$.
The other two properties, instead, could be obtained observing that the bump in \figurename~\ref{fig:binderBump3D} occurs before $d=1$ and that for $d=1$ the dependence of $\Binder\bigl[\mathcal{P}(x)\bigr]$ on $\alpha$ drops out,
\begin{equation}\label{eq:Binderd1}
    \Binder\bigl[\mathcal{P}(x)\bigr]_{d=1}=3-\frac{6}{(2+3\sigma^2)^2}\;.
\end{equation}
Then one could choose the function $d(\beta)$ such that $d(\beta_c)=1$ and choose $\sigma$ in order to have the desired value of the Binder cumulant at the critical temperature.
For the case of interest, i.e. when the \RW{} endpoint is a triple point and $\Binder=1.5$, one should choose in our simple model $\sigma=0$, which is clearly not allowed on finite volumes.
Nevertheless, the standard deviation is known to go to 0 in the thermodynamic limit, when the Binder cumulant takes the universal value.
We will come back to this aspect later in the section.
For the moment, if we just decide to reproduce our data, we have to set $\Binder$ to the measured value, that is usually higher than the theoretical one (as observed in~\cite{deForcrand:2010he,Philipsen:2014rpa}).
For example, in \figurename{}s~\ref{fig:binderBump2D}, \ref{fig:distribution}, and \ref{fig:binderBump3D} we fixed $\sigma=0.1$ that would mean $\Binder(\beta_c)\simeq 1.544$ only slightly higher than 1.5.
Instead, the value $\Binder(\beta_c)=1.68$ extracted from our data at $\kappa=0.165$ would lead to a not so large $\sigma\simeq0.21$, yet larger than suggested by the actual data.
Another property that the function $d(\beta)$ should reproduce is the fact that for larger $\NSigma$ the transition happens faster.
We already noticed that $\alpha$ reproduces this feature in our model.
Hence it makes sense to assume $\alpha\:\propto\:\NSigma$ and to let $d$ depend also on $\alpha$.
As function of $\beta$, $d(\alpha,\beta)$ has to change more drastically around $\beta_c$ for increasing values of $\alpha$.
One possibility which also fulfills the requirements for $\beta\to 0$ and for $\beta=\beta_c$ is
\[
    d(\alpha,\beta)=\frac{e^{\,\alpha\beta}-1}{e^{\,\alpha\beta_c}-1}\;.
\]
Inserting this choice in the expression of $\Binder\bigl[\mathcal{P}(x)\bigr]$, it is possible to plot the Binder cumulant as function of $\beta$ for fixed $\sigma=0.1$ and for some values of $\alpha$ (that plays the role of $\NSigma$).
This has been done in \figurename~\ref{fig:binderBump2D}.
The similarity to \figurename~\ref{fig:k1650bump} is evident.
In particular, in both figures the bump shrinks and its height grows as the volume is increased.
Naturally, it is also possible to take the thermodynamic limit, that means let $\alpha\to\infty$.
To do that it is sufficient to notice that
\[
    \lim_{\alpha\to\infty}\: \alpha^m \bigl[d(\alpha,\beta)\bigr]^n=
    \lim_{\alpha\to\infty}\: \alpha^m e^{n\alpha(\beta-\beta_c)}=
    \left\{
    \begin{aligned}
        0      &\;,\; \beta<\beta_c \\
        \infty &\;,\; \beta>\beta_c
    \end{aligned}
    \right.
\]
for integers $n>0$ and $m\geq0$. Using this relation in the expression of the Binder cumulant we get
\begin{equation}\label{eq:BinderThermLimit}
    \lim_{\alpha\to\infty} \;\Binder\bigl[\mathcal{P}(x)\bigr]=
    \left\{
    \begin{aligned}
        3 &\quad\text{for}\quad \beta<\beta_c   \\
        1 &\quad\text{for}\quad \beta>\beta_c
    \end{aligned}
    \right.\;,
\end{equation}
which is exactly the expected behaviour in the thermodynamic limit.
At $\beta=\beta_c$ we already showed in \Eq{eq:Binderd1} that the Binder cumulant does not depend on $\alpha$ and that fixing $\sigma$ to some finite, small value brings it to $\Binder>1.5$, i.e. not exactly the universal value.
Nevertheless, it is sufficient to assume $\sigma\:\propto\:\alpha^{-1}$ to completely reproduce the physical situation.
In particular, this means that the standard deviation goes to 0 for $\alpha\to\infty$, which implies
\[
    \lim_{\alpha\to\infty} \;\Binder\bigl[\mathcal{P}(x)\bigr]_{\beta=\beta_c}=1.5\;
\]
(observe how the limits in \Eq{eq:BinderThermLimit} are still valid assuming $\sigma$ proportional to $\alpha^{-1}$).
The Binder cumulant bump is then nothing but a finite size effect!
This suggests that also the larger than expected value $\Binder(\beta_c,\infty)$ is due to these corrections.


\section{Numerical results and discussion}\label{sec:results}

\begin{figure*}
    \centering
    \subfigure[ $\kappa=0.1$, first-order coefficients.]
    {\label{fig:collapseA}\includegraphics[width=.45\textwidth]{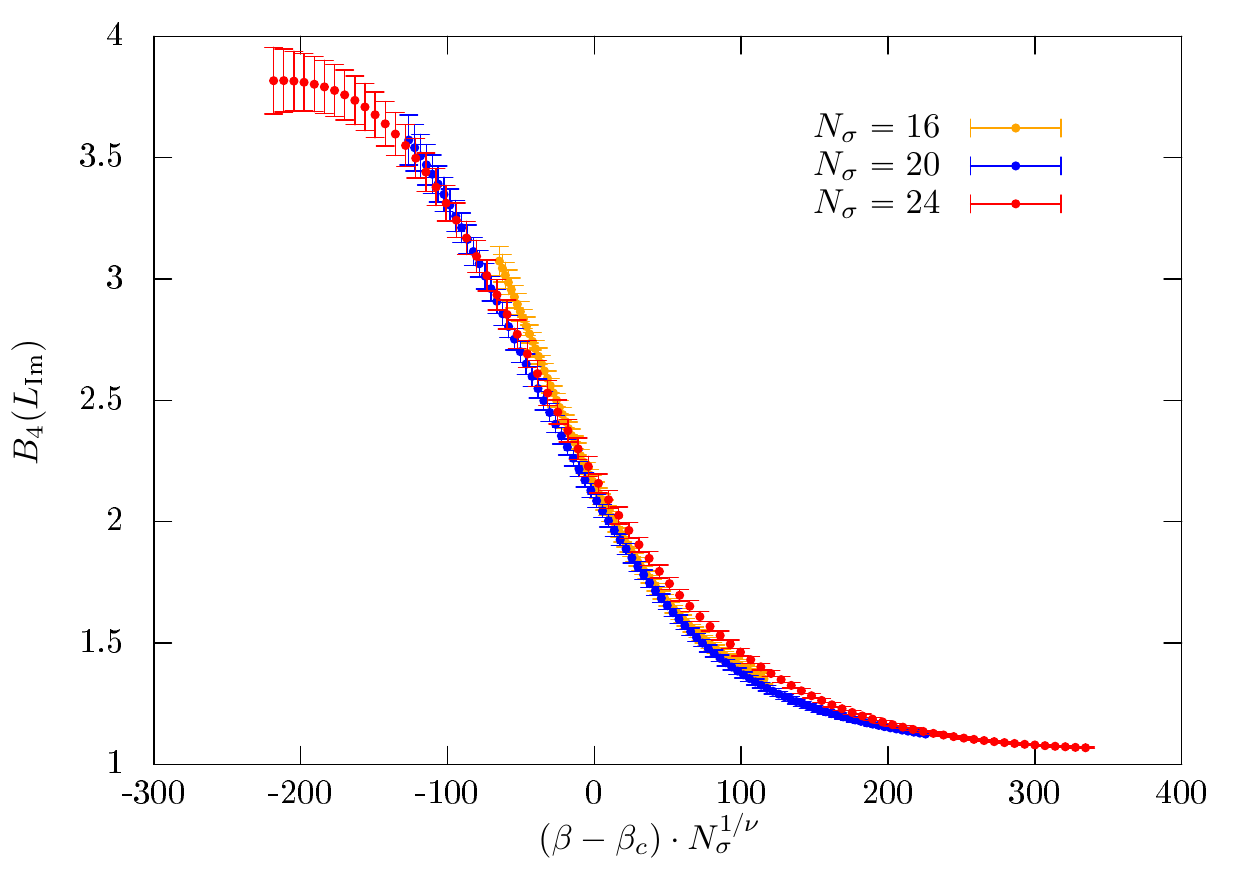}} \qquad
    \subfigure[ $\kappa=0.1$, second-order coefficients.]
    {\label{fig:collapseB}\includegraphics[width=.45\textwidth]{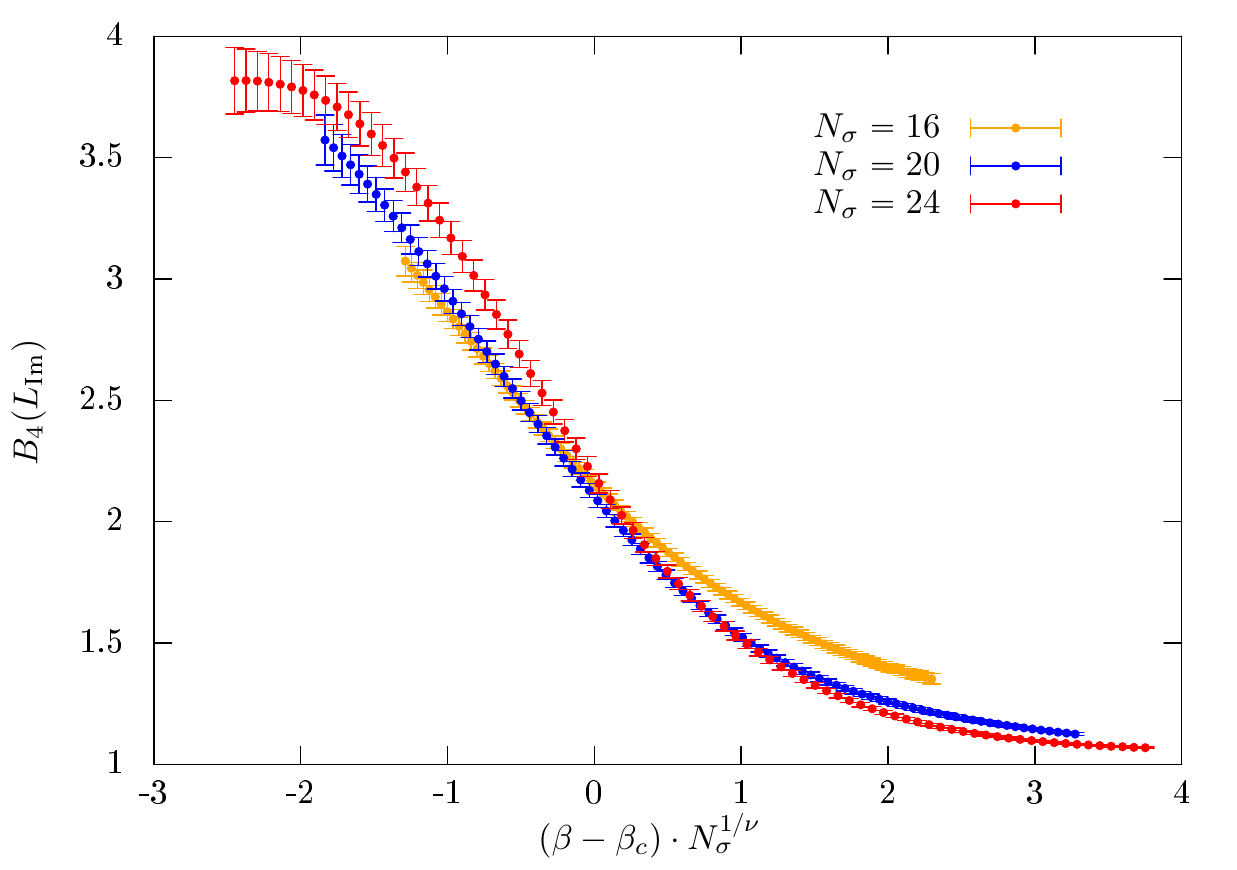}} \\
    \subfigure[ $\kappa=0.13$, first-order coefficients.]
    {\label{fig:collapseC}\includegraphics[width=.45\textwidth]{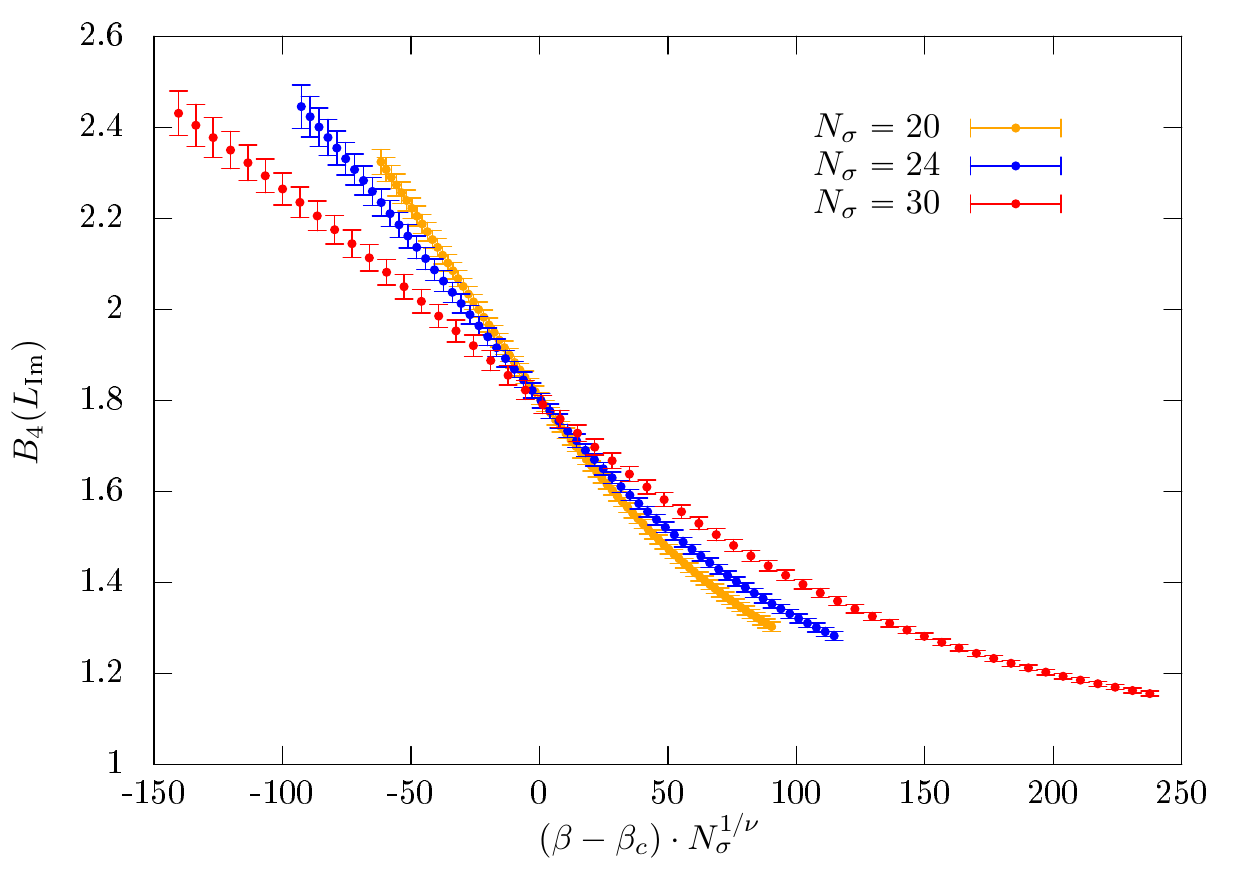}} \qquad
    \subfigure[ $\kappa=0.13$, second-order coefficients.]
    {\label{fig:collapseD}\includegraphics[width=.45\textwidth]{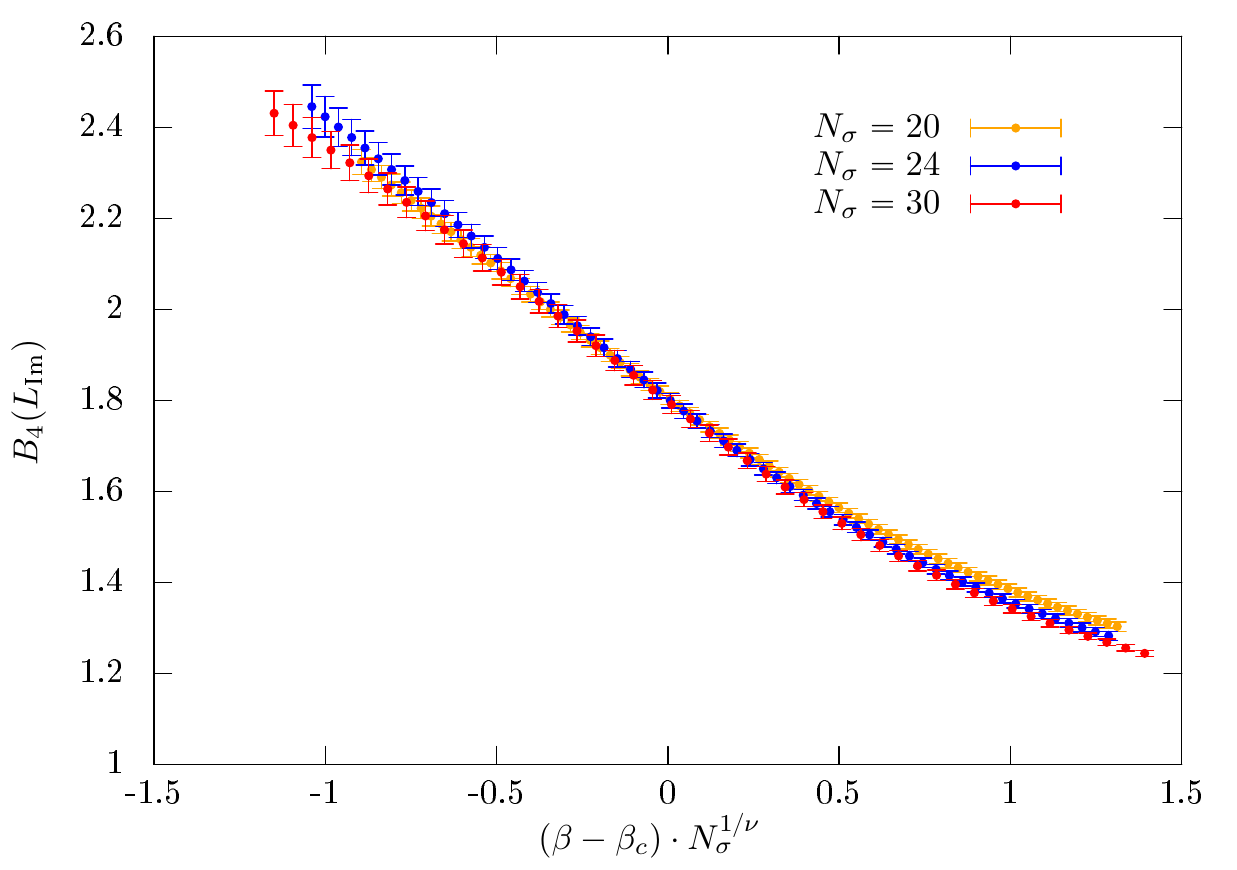}} \\
    \subfigure[ $\kappa=0.165$, first-order coefficients.]
    {\label{fig:collapseE}\includegraphics[width=.45\textwidth]{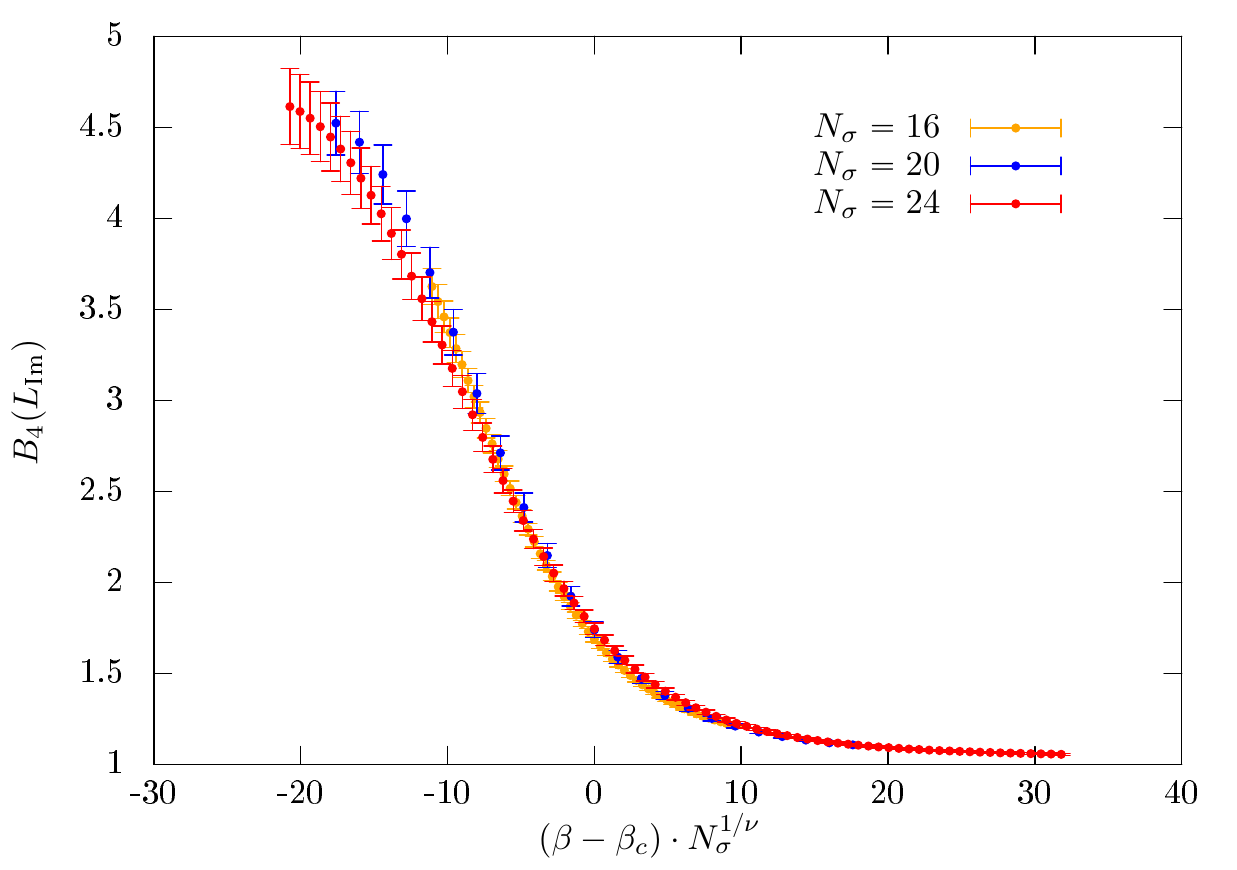}} \qquad
    \subfigure[ $\kappa=0.165$, second-order coefficients.]
    {\label{fig:collapseF}\includegraphics[width=.45\textwidth]{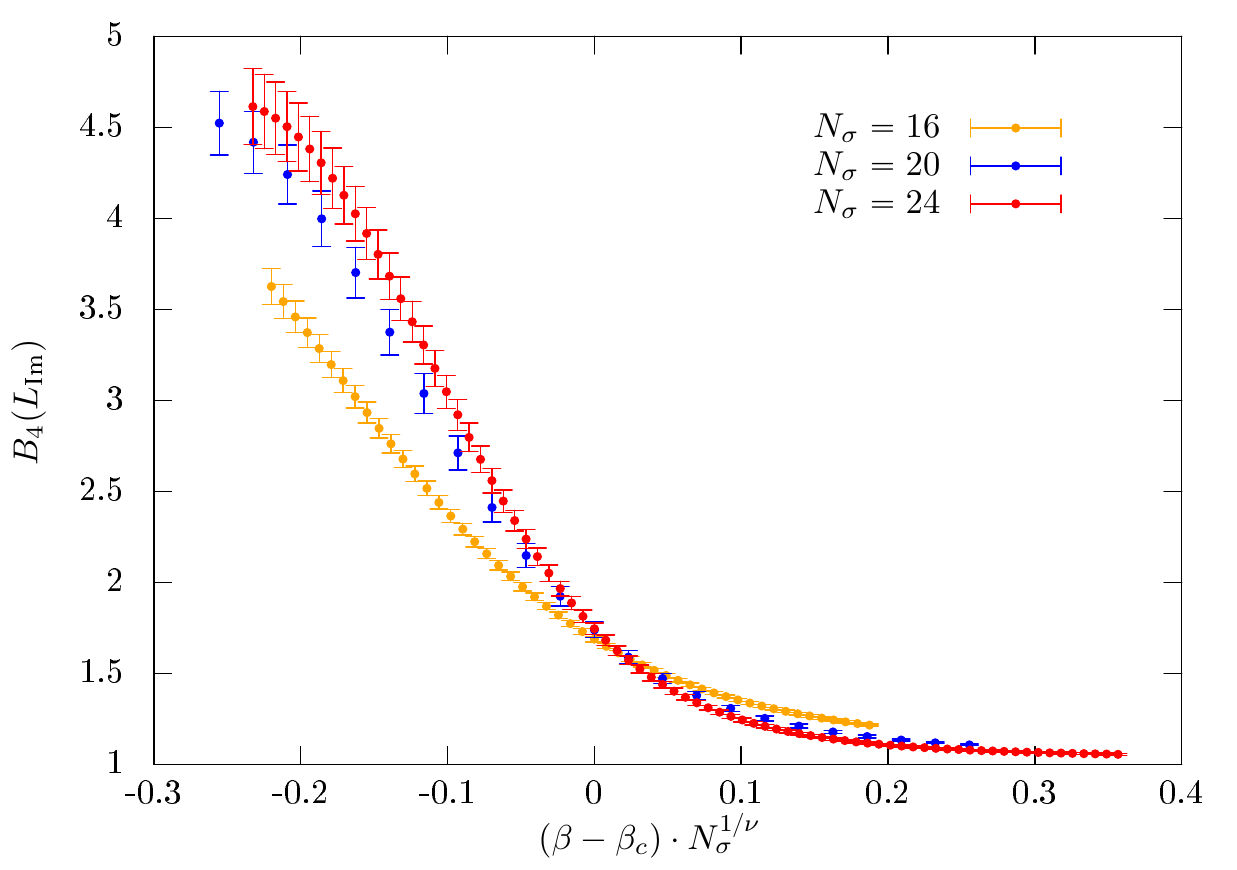}}
    \caption{Example of collapse plots of the Binder cumulant of the imaginary part of the \PL.}
    \label{fig:collapseBinder}
\end{figure*}

To get a first impression about the nature of the phase transition, we produced collapse plots of the susceptibilities 
at each value of $\kappa$ according to \Eq{eq:SuscScaling}, where the norm $\lVert\Poly\rVert$ of the \PL{} was used as observable.
Because of the different numerical values of the ratios $\gamma/\nu$ for a first- and a second-order phase transition, the collapse plots usually help to exclude one scenario.
However, especially for low $\NSigma$, the collapse plots of the susceptibilities are often inconclusive and we complement them with collapse plots of the Binder cumulant of the imaginary part of the \PL{} according to \Eq{eq:BinderFSS}.
In \figurename~\ref{fig:collapseBinder}, we show examples at $\kappa=0.1, \kappa=0.13$ and $\kappa=0.165$ with first-order exponents in the left column and second-order exponents in the right column.
In each case, the quality of the collapse clearly prefers one set of critical exponents. This indicates that 
$\kappa=0.1$ and $\kappa=0.165$ are in the first-order region, while $\kappa=0.13$ is in the second-order region.
Note how the Binder cumulant takes values larger than 3 for the first-order $\kappa$, as discussed in the previous section, while it does not for the intermediate ones.

\begin{figure*}
    \centering
    \subfigure[ $\NTau=4$, result of \referencename~\onlinecite{Philipsen:2014rpa}.]
    {\label{fig:nuVSkappaA}\includegraphics[width=.45\textwidth]{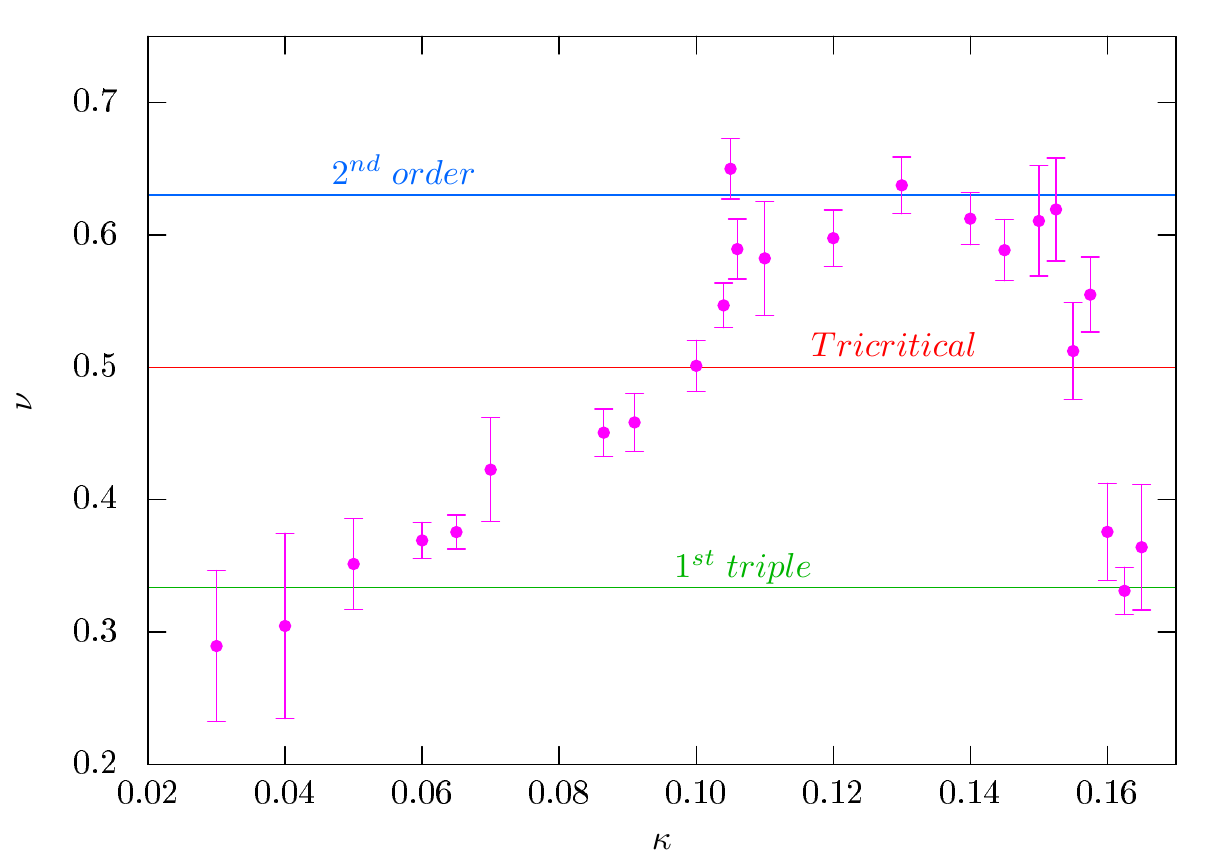}} \qquad
    \subfigure[ $\NTau=6$, data of present work.]
    {\label{fig:nuVSkappaB}\includegraphics[width=.45\textwidth]{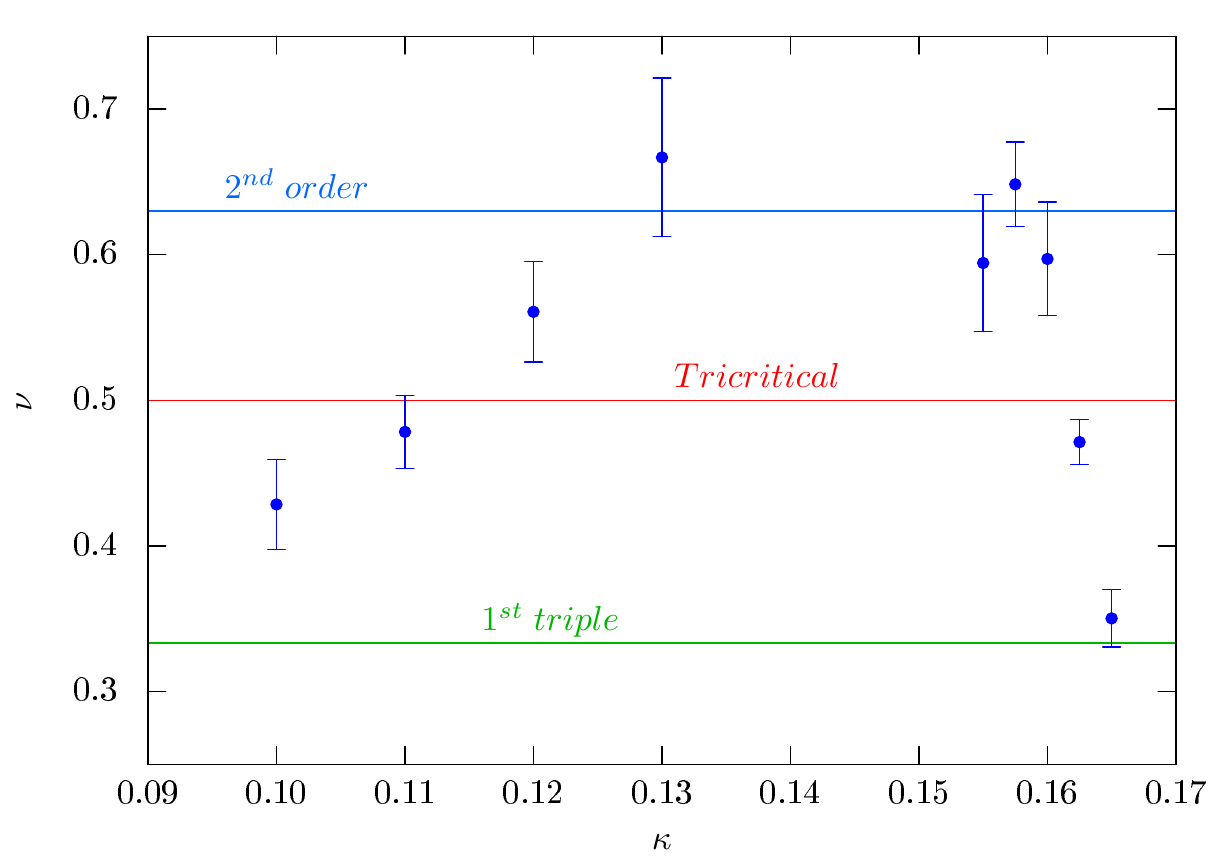}}
    \caption{Critical exponent $\nu$ as function of $\kappa$. The horizontal coloured lines are the critical
             values of $\nu$ for some universality classes.
             Note the different scale on the $\kappa$-axis.
             Due to the much higher numerical cost, not all the $\kappa$ values simulated for $\NTau=4$ have
             been considered for $\NTau=6$. 
             Refer to \figurename~\ref{fig:nuVSkappaComparisonA} for a more direct comparison.}
    \label{fig:nuVSkappa}
\end{figure*}

The collapse plot technique is useful as an orientation, but it is only self-consistent and we also wish to
actually calculate the critical exponents. Thus we fit the Binder cumulant data to \Eq{eq:BinderFSS}, obtaining the critical exponent $\nu$ as a fit parameter.
In order to have objective fitting criteria and avoid ``fits by eye'' we developed an intricate procedure which is detailed in \appendixname~\ref{app:nu}.
\figurename~\ref{fig:nuVSkappa} shows the values of $\nu$ extracted from the fits, plotted as function of $\kappa$.
As expected, $\nu$ changes from first- to second-order values and back again.
This behaviour approaches a step function in the thermodynamic limit but remains smoothed out when the lattice volume is finite.
In particular, this means that $\nu$ can in principle take any value between the universal ones in the 
crossing region, while
far away from the tricritical masses, it is compatible with $1/3$ (first order) for small and large $\kappa$, and with $0.6301(4)$ (second order) for intermediate $\kappa$.
From the fit, the value of the Binder cumulant at the critical coupling in the infinite volume limit, $\Binder(\beta_c,\infty)$, can be extracted as well.
In agreement with previous studies both with staggered fermions~\cite{deForcrand:2010he} and with Wilson fermions~\cite{Philipsen:2014rpa}, this value is slightly higher than the universal one, due to finite volume corrections as discussed in \sectionname~\ref{sec:bump}.
However, the critical exponent $\nu$ suffers much less from this problem and is well suited to understand the nature of the phase transition. 
In accordance with these expectations, we estimate the two tricritical values of $\kappa$ as
\begin{equation}
    \LatMassWilsonTricHeavy=0.11(1) ,\quad  \LatMassWilsonTricLight=0.1625(25) \:.
\end{equation}

\begin{figure*}
  \centering
  \subfigure[$\NTau=4$ and $\NTau=6$ results.]
  {\label{fig:nuVSkappaComparisonA}\includegraphics[width=.45\textwidth]{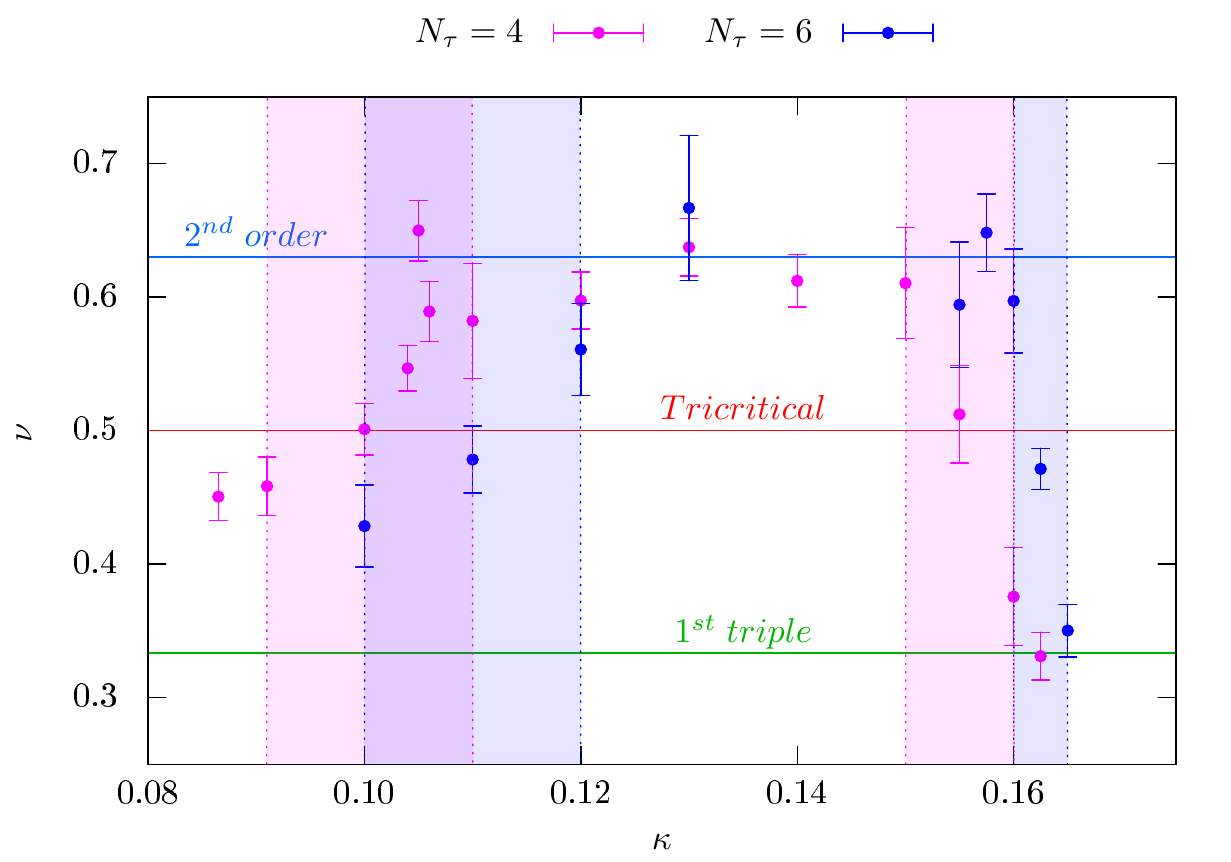}} \qquad
  \subfigure[Results in terms of \mpi.]
  {\label{fig:nuVSmassComparisonB}\includegraphics[width=.45\textwidth]{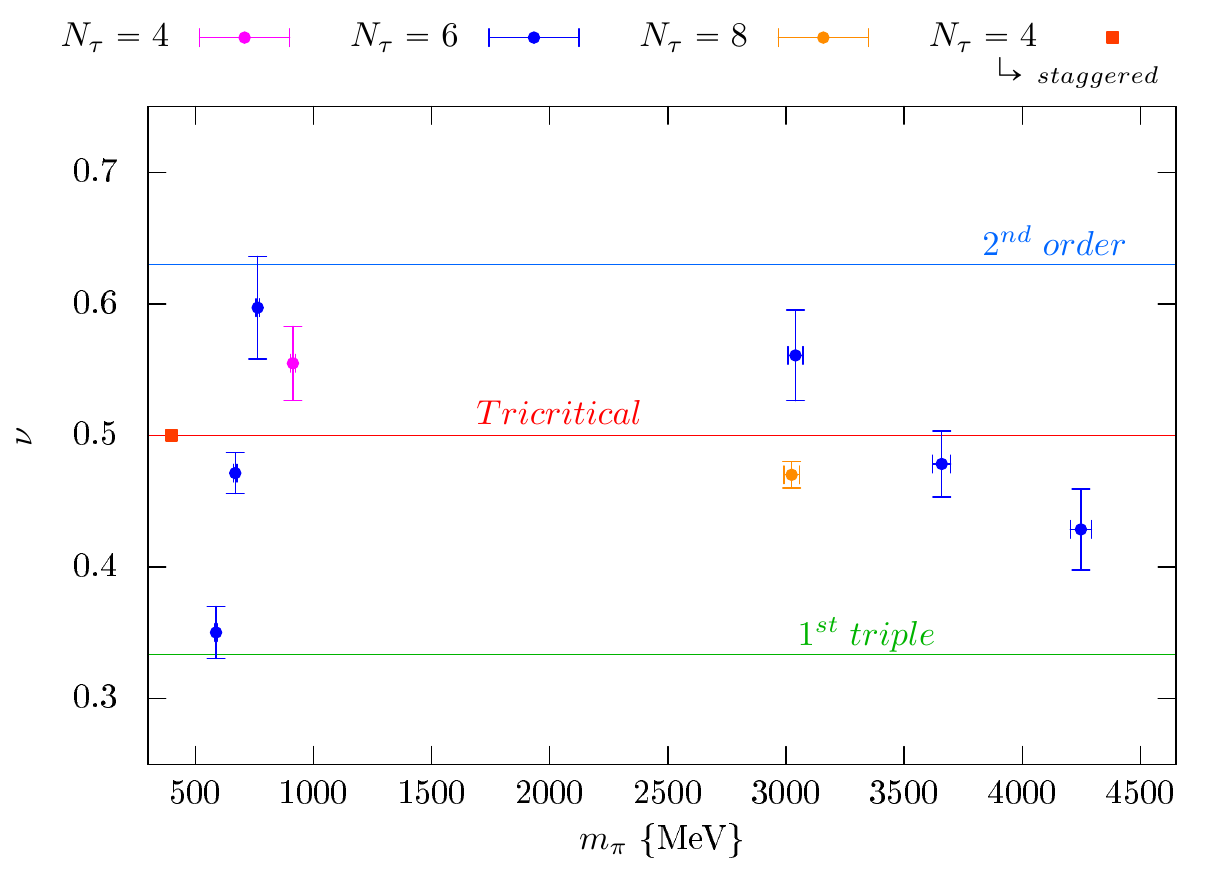}}
  \caption{Direct comparison between $\NTau=4$ and $\NTau=6$ results and comparison of $\NTau=4,6,8$ results in terms of \mpi.
           In the latter case, the value of \LatMassWilsonTricLight\ from \cite{Bonati:2010gi} has been included as well.
           For the sake of clarity, not all the $\NTau=4$ points have been included.
           The vertical coloured bands highlight the position of the tricritical masses.
           A shift toward small masses (i.e. bigger $\kappa$) is evident as $\NTau$ is increased.}
\end{figure*}

\newcolumntype{L}{>{\centering\arraybackslash}p{0.2\columnwidth}}
\newcolumntype{J}{>{\centering\arraybackslash}p{0.16\columnwidth}}
\newcolumntype{K}{>{\centering\arraybackslash}p{0.08\columnwidth}}
\newcolumntype{?}{!{\vrule width 0.3mm}}
\renewcommand{\arraystretch}{1.2}
\begin{table*}[ht]
  \centering
  \begin{tabular}{JJJ@{\quad}|@{\quad}S[table-format=3.8]S[table-format=3.8]@{\quad}|@{\quad}LL@{\quad}?@{\quad}KL}
    \toprule[0.3mm]
    $\kappa$ & $\LatCoupling$ & \texttt{\# confs} & {$w_0/a$} & {$a\,\mpi$} & $\LatSpacing\!$ \{fm\} & \mpi \{MeV\} & $\NTau$ & $T\!$ \{MeV\} \\
    \midrule[0.1mm] 
    0.0910 & 5.6655 & 1600 & 0.9161(6)  & 3.0107(2)  & 0.192(2) & 3101(32) & \multirow{4}{*}{4} & 258(3)     \\
    0.1000 & 5.6539 & 1600 & 0.9017(12) & 2.7285(2)  & 0.195(2) & 2766(29) &                    & 253(3)      \\
    0.1100 & 5.6341 & 1600 & 0.8789(10) & 2.4250(3)  & 0.200(2) & 2396(25) &                    & 247(3)      \\
    0.1575 & 5.3550 &  400 & 0.7104(3)  & 1.1426(17) & 0.247(3) & 913(9)   &                    & 200(2)      \\
    \midrule[0.1mm]
    0.1000 & 5.8698 & 1600 & 1.4650(20) & 2.5793(6)  & 0.120(1) & 4248(44) & \multirow{6}{*}{6} & 275(3)       \\
    0.1100 & 5.8567 & 1600 & 1.4594(18) & 2.2302(2)  & 0.120(1) & 3659(38) &                    & 273(3)      \\
    0.1200 & 5.8287 & 1200 & 1.4333(20) & 1.8862(4)  & 0.122(1) & 3040(31) &                    & 269(3)      \\
    0.1600 & 5.4367 &  200 & 1.1248(14) & 0.6045(15) & 0.156(2) & 764(8)   &                    & 211(2)      \\ 
    0.1625 & 5.3862 &  200 & 1.0700(17) & 0.5559(23) & 0.164(2) & 669(8)   &                    & 201(2)      \\ 
    0.1650 & 5.3347 &  200 & 1.0082(13) & 0.5184(27) & 0.174(2) & 588(7)   &                    & 189(2)      \\ 
    \midrule[0.1mm]
    0.1300 & 5.9590 & 1600 & 1.9357(44) & 1.3896(2)  & 0.091(1) & 3024(32) &          8         & 272(3)        \\ 
    \bottomrule[0.3mm] 
  \end{tabular}
  \caption{Results of the scale setting ($T=0$ simulations performed on $\NTau=32$, $\NSigma=16$ lattices).
           The number of independent configurations used is reported in the third column (\texttt{\# confs}).
           $w_0/a$ has been determined and converted to physical scales using the publicly available code described in \referencename~\onlinecite{Borsanyi:2012zs}.
           For the pion mass determination, eight point sources per configuration have been used.
           The table also contains the lattice spacing, the pion mass and the temperature of the corresponding finite temperature ensemble in physical units.}
  \label{tab:scaleSetting}
\end{table*}

For comparison, the results from $\NTau=4$ \cite{Philipsen:2014rpa} are also shown in Figure \ref{fig:nuVSkappaComparisonA}. In accord with expectations, both tricritical (bare) masses 
move to smaller values on the finer lattice. 
To convert these findings into universal and physical units, we set the scale at or close to the respective \LatCouplingC\ for the relevant $\kappa$.
The results for the lattice spacing \LatSpacing, the critical temperature \Tc\ and \mpi\ are summarized in Table \ref{tab:scaleSetting}.
Since the scale setting method using $w_0$ is much more precise than using the $\rho$ mass as in \referencename~\onlinecite{Philipsen:2014rpa}, we evaluated again the $\Temp=0$ simulations from the latter study and include them here for completeness.
In addition, we performed $T=0$ simulations for the $\NTau=4$ \LatMassWilsonTricHeavy\ values.
The lattices coarsen going to lower masses, since \LatCoupling\ decreases.
All lattices considered are coarse, $0.12 $ fm $ \lesssim\LatSpacing\lesssim 0.18$ fm.
However, compared to the $\NTau=4$ simulations, where $\LatSpacing \gtrsim 0.19$ fm, a clear decrease in \LatSpacing\ is achieved, as expected.
Note that $\mpi\,L > 6$ for all our parameter sets, so that finite size effects are negligible.

Our estimates of the tricritical points in physical units for the given lattice spacing then read
\begin{align*}
    m_\pi^\text{tricr. heavy}&=3659^{+589}_{-619}\text{ MeV} \;,\\[1ex]
    m_\pi^\text{tricr. light}&=669^{+95}_{-81}\text{ MeV} \;.
\end{align*}
Note that the heavy masses in lattice units are much larger than one.
Hence the continuum mass estimates still suffer from large cut-off effects.
Thus, the quoted number for  $m_\pi^\text{tricr. heavy}$ still contains a large systematic error and a quantitative evaluation of its shift from coarser lattices is impossible.
On the other hand, the shift in the lower tricritical mass is from
$\mpi \approx 910$~MeV to $\mpi \approx 670$~MeV, or around 35\%.
By contrast, the critical temperature \Tc\ does not seem to depend much on $\NTau$ and stays roughly constant at around $200$ MeV.

Our shifts in the tricritical pion masses are of similar magnitude as those in the $\Nf=3$ critical pion masses 
at $\mu=0$ with Wilson Clover fermions \citep{Jin:2014hea}.
Comparing our results to Ref. \cite{Bonati:2010gi}, one sees that our lighter tricritical mass on $\NTau=6$
is still higher than the staggered estimate from $\NTau=4$, which is roughly $400$ MeV.
Altogether this shows that $\NTau\leq 6$ is still far from the region where linear cut-off effects dominate
in the standard Wilson action and suggests that drastically larger $\NTau$ are required for both discretizations.
This is expected from studies of the equation of state, where different discretization start to agree at $\NTau \gtrsim 12$ only (see Ref. \onlinecite{Philipsen:2012nu} for a recent overview).

As a first step towards larger $\NTau$, we also performed simulations at $\NTau=8$ and $\kappa=0.13$, with $\NSigma = 16,24,32,40$, corresponding to aspect ratios of $2-5$, (for details, see~\appendixname~\ref{app:sim}).
The computational costs increase dramatically with \NTau{} and the statistics gathered for the $\NSigma=40$ simulations is not as high as for the previous simulations.
However, $\nu$ can be determined in a solid fashion using the data for the other three spatial volumes, giving a value of $\nu = 0.47(1)$.
The lattice spacing \LatSpacing\ is now reduced from $\approx 0.12$~fm to $\approx 0.09$~fm.
In physical units, this new point is located at $\mpi=3024(32)$.
Given the same caveats discussed for $\NTau=6$, this again suggests a large shift for the heavy tricritical mass.
Note that \Tc\ stays again constant when going from $\NTau=6$ to $8$.
Our findings are summarized in Figure \ref{fig:nuVSmassComparisonB}, that compares the tricritical regions for the different \NTau.
Also included is the $\NTau=4$ value from staggered studies \cite{Bonati:2010gi}.
The figure makes apparent that much larger \NTau\ are required in order to go to the continuum.

\section{Conclusions}
\label{sec:summary}

We have extended previous studies of the nature of the Roberge-Weiss endpoint of $\Nf=2$ QCD 
at imaginary chemical potential
to $\NTau=6$ and for one mass value to $N_\tau=8$, using standard Wilson fermions. 
To this end, we gathered large amounts of data for several volumes and carried out a thorough  
finite size analysis. In particular, we have understood the occurrence of a ``bump'' in the 
Binder cumulant in the region where the Roberge-Weiss endpoint is a triple point. The behaviour can 
be explained as a finite size effect specifically due to the merging of a three peak distribution to a two peak
distribution as a function of the lattice coupling.

The qualitative phase structure fully replicates that on the coarser $\NTau=4$ lattices.
However, the tricritical pion mass values separating the regime of a second-order endpoint from triple points
in the small and large mass region shift considerably when the cut-off is reduced and suggest that
significantly finer lattices are necessary before the observed phase structure settles quantitatively in 
the continuum.


\begin{acknowledgments}
We thank the staff of \Loewe\ and \Lcsc\ for its support, Andrei Alexandru for early discussions and Frederik Depta
for code to extract the pion mass.

C.C, O. P., C. P. and A.S. are supported by the Helmholtz International Center for FAIR within the LOEWE program of the State of Hesse.
C.C. is supported by the GSI Helmholtzzentrum f\"ur Schwerionenforschung.
F.C. and O.P. are supported by the German BMBF under contract no. 05P1RFCA1/05P2015 (BMBF-FSP 202). 

\end{acknowledgments}

%


\onecolumngrid

\appendix

\section{Simulation details}\label{app:sim}
\begin{figure*}[t]
    \centering
    {\includegraphics[width=.47\textwidth]{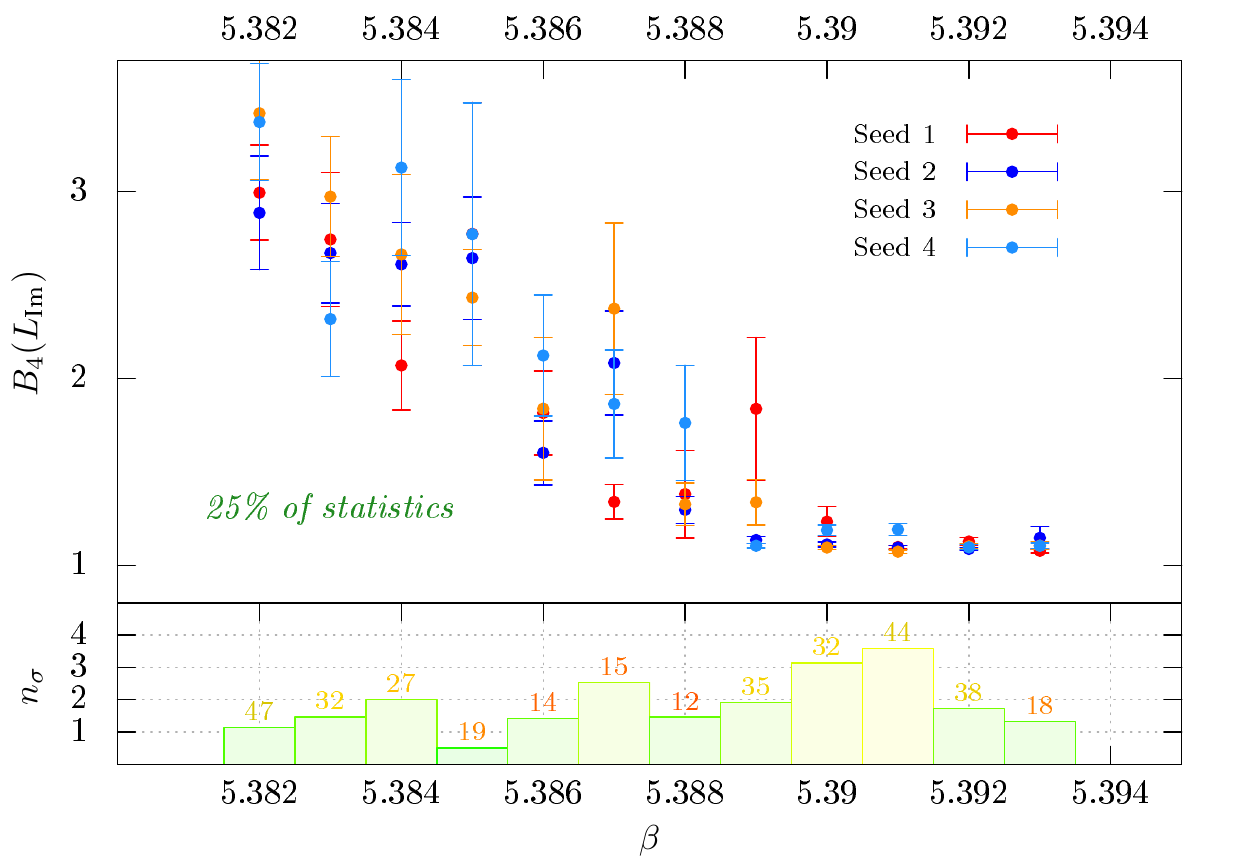}} \qquad
    {\includegraphics[width=.47\textwidth]{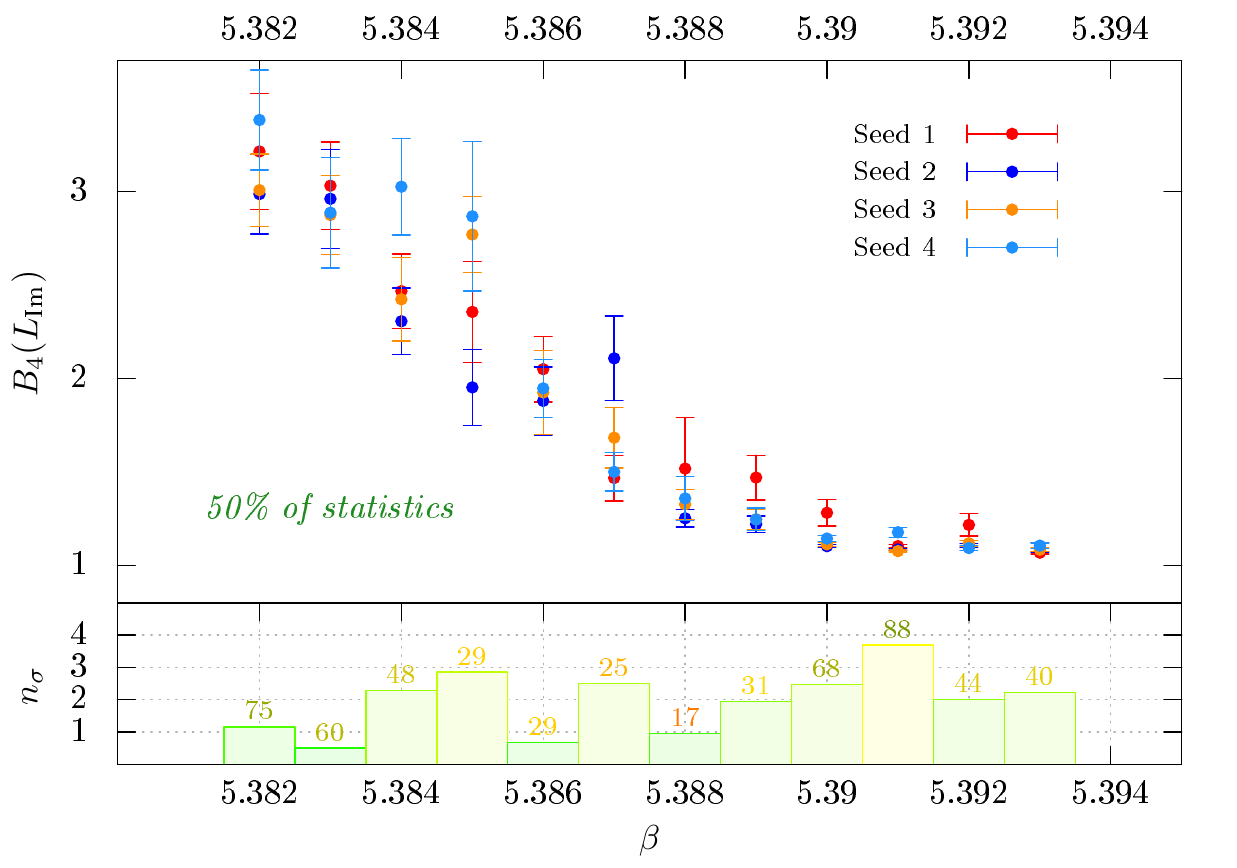}} \\
    {\includegraphics[width=.47\textwidth]{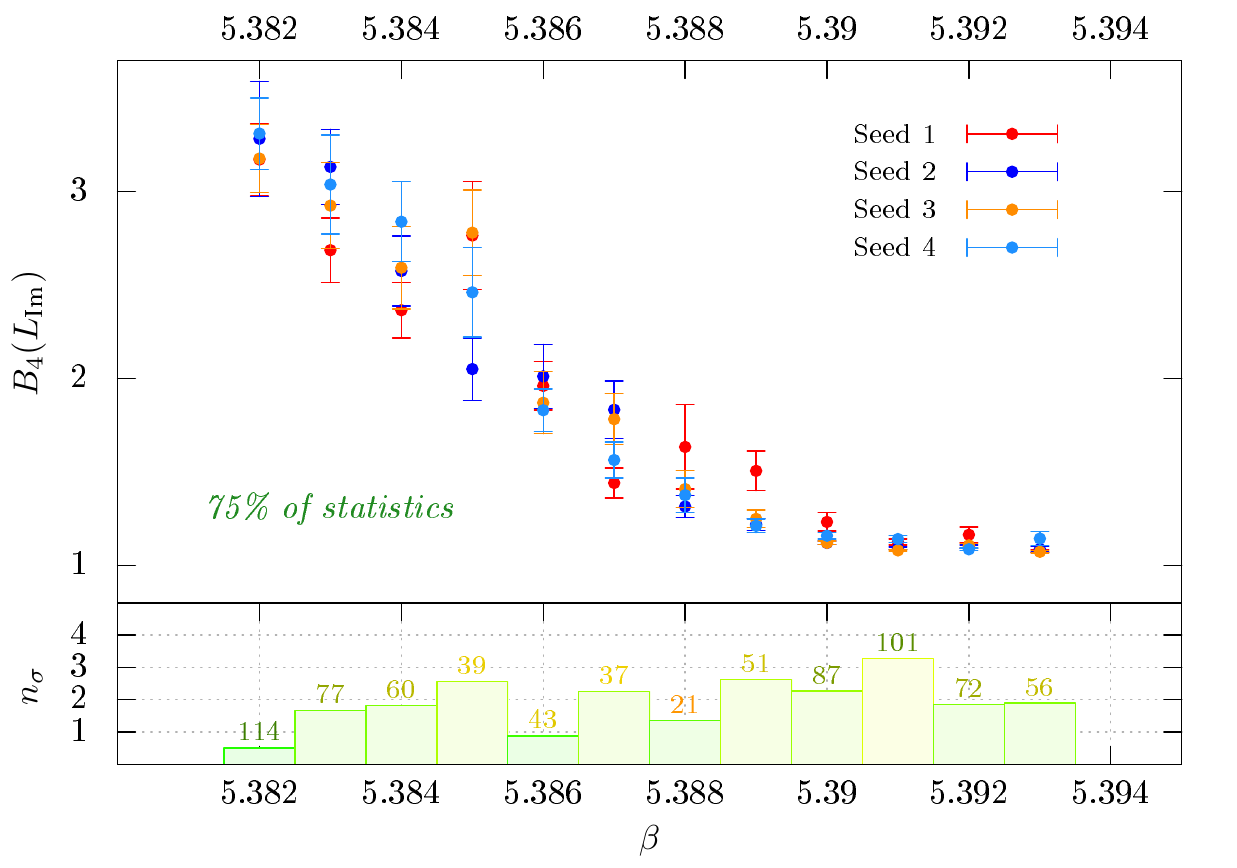}} \qquad
    {\includegraphics[width=.47\textwidth]{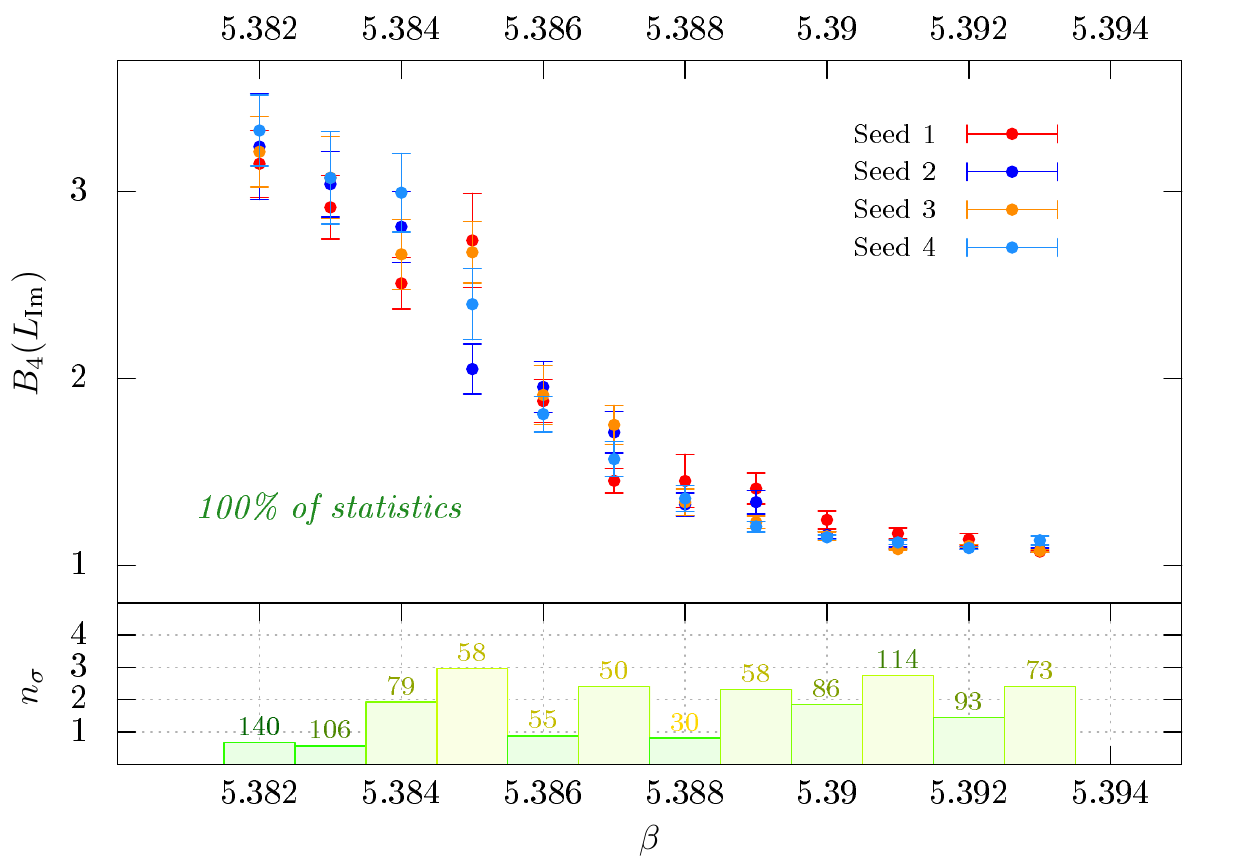}}
    \caption{Successive analysis of the Binder cumulant measurements at $\kappa=0.1625$ on $\NSigma=18$.
             The histogram below each plot is a guideline to judge on the statistics. $n_\sigma$ at each
             $\beta$ is the number of standard deviations at which the two most different chains are compatible.
             The number above each bar is the average number of independent events collected at that $\beta$.
             The colors have been chosen in order to reflect the goodness of the statistics: from green
             (statistics high enough) to red (statistics to be increased). Both $n_\sigma$ and the number
             of independent events have to be monitored to decide when to stop increasing statistics.}
    \label{fig:chains}
\end{figure*}

A detailed overview of all our simulation runs is provided in \tablename~\ref{tab:simOverview}.
Measurements of the Binder cumulant are difficult because of the large autocorrelations involved and 
the large statistics required. 
For a generic observable $x$, the sets of measurements $\{x_i\}$, $\{x_i^2\}$, \ldots, $\{x_i^n\}$ show different integrated autocorrelation times \TauInt{}, which we estimate using the Wolff algorithm \cite{Wolff:2003sm}. 
Dividing the total number of HMC trajectories by \TauInt{} gives the number of independent measurements for a  given observable. We collected at least 30 independent events per run of a given parameter set 
for $\Binder(\PolyIm)$.
In addition, we run the same parameter set generating typically four independent Markov chains
until $\Binder(\PolyIm)$  is compatible within three standard deviations between all of them.  
\figurename~\ref{fig:chains} shows an example at $\kappa=0.1625$ on $\NSigma=18$.
The improvement of the signal with statistics is clearly visible.
Once each chain is long enough, we merged them for the finite size scaling analysis.

\DeclareExpandableDocumentCommand{\cc}{O{gray!15} m}{\multicolumn{1}{>{\columncolor{#1}[1.01\tabcolsep][1.01\tabcolsep]}c}{#2}} 
\newcolumntype{P}{>{\centering\arraybackslash}p{0.13\textwidth}}
\newcolumntype{Q}{>{\centering\arraybackslash}p{0.04\textwidth}}
\newcolumntype{R}{>{\centering\arraybackslash}p{0.08\textwidth}}
\renewcommand{\arraystretch}{1.2}
\def\colorXVI{Cyan!20}
\def\colorXVIII{Magenta!15}
\def\colorXXX{Goldenrod!20}
\def\colorXXXII{Green!20}
\def\colorXII{red!20}
\def\colorXXXVI{Orange!20}
\def\colorXL{Violet!15}

\begin{table}[b]
  \centering
  \begin{tabularx}{\textwidth}{QRP@{\qquad}*{5}{P}}
    \toprule[0.3mm]
    \multirow{2}{*}{\NTau} & \multirow{2}{*}{$\kappa$} & \multirow{2}{*}{$\beta$ range} &
    \multicolumn{5}{c}{Total statistics per spatial lattice size \NSigma{} $\bigl($ \texttt{\#} of simulated $\beta$ values | \texttt{\#} of chains$\bigr)$} \\
    & & & \colorbox{\colorXVI}{16} \colorbox{\colorXVIII}{18} & 20 & 24 & \colorbox{\colorXXX}{30} \colorbox{\colorXXXII}{32}
    & \colorbox{\colorXII}{12} \colorbox{\colorXXXVI}{36} \colorbox{\colorXL}{40} \\
    \midrule[0.1mm]
    \multirow{9}{*}{6}
    & 0.1000 & 5.8460-5.9020 & \cc[\colorXVI]{6.11M (24 | 2)}    & 4.36M (16 | 2) & 4.30M (16 | 2) &               -                &-\\
    & 0.1100 & 5.8400-5.8660 &               -                   & 3.81M (26 | 4) & 1.49M (14 | 4) & \cc[\colorXXX]{4.05M (18 | 4)} & \cc[\colorXXXVI]{1.92M (13 | 4)}\\
    & 0.1200 & 5.8180-5.8450 & \cc[\colorXVI]{5.28M (10 | 4)}    & 3.89M ( 9 | 4) & 3.23M ( 9 | 4) & \cc[\colorXXX]{2.19M ( 8 | 4)} &-\\
    & 0.1300 & 5.7760-5.7980 &               -                   & 3.94M (25 | 4) & 3.76M (23 | 4) & \cc[\colorXXX]{3.56M (16 | 4)} &-\\
    & 0.1550 & 5.5210-5.5420 & \cc[\colorXVI]{1.40M (30 | 1)}    & 1.04M (23 | 1) & 1.12M (24 | 1) & \cc[\colorXXX]{0.76M ( 9 | 4)} &-\\
    & 0.1575 & 5.4750-5.4930 & \cc[\colorXVIII]{0.59M ( 7 | 4)}  &        -       & 0.92M ( 7 | 4) & \cc[\colorXXX]{1.40M ( 7 | 4)} &-\\
    & 0.1600 & 5.4330-5.4430 & \cc[\colorXVIII]{0.52M ( 6 | 4)}  &        -       & 0.86M ( 6 | 4) & \cc[\colorXXX]{1.12M ( 6 | 4)} &-\\
    & 0.1625 & 5.3800-5.3930 & \cc[\colorXVIII]{0.92M (12 | 4)}  &        -       & 1.12M ( 8 | 4) &               -                & \cc[\colorXII]{1.38M (7 | 4)}\\
    & 0.1650 & 5.3260-5.3370 & \cc[\colorXVI]{1.99M (16 | 4)}    & 1.09M (11 | 4) & 1.71M (12 | 4) &               -                &-\\
    \midrule[0.1mm] 
    \multirow{1}{*}{8} 
    & 0.1300 & 5.9400-5.9800 & \cc[\colorXVIII]{3.69M (9 | 4)}  &        -       & 5.40M (9 | 4) & \cc[\colorXXXII]{2.00M (5 | 4)} &    \cc[\colorXL]{1.00M (5 | 4)} \\
    \bottomrule[0.3mm] 
  \end{tabularx}
  \caption{Overview of the statistics accumulated in all the simulations ($\NTau=6$ and $\MuI=\pi/6$).
           Since the resolution in $\beta$ is not the same at different $\kappa$, the number of simulated
           $\beta$ has been reported per each range. The accumulated statistics per $\beta$ has not always 
           been the same. Therefore the number of trajectories here is about all the trajectories produced per
           given \NSigma. Using the number of chains provided above, it can be easily estimated how long
           was on average each chain, even though we always accumulated higher statistics close to the
           critical temperature.}
  \label{tab:simOverview}
\end{table}

\section{Extracting the critical exponent \texorpdfstring{$\mathbb{\nu}$}{nu}}\label{app:nu}
As described in \sectionname~\ref{sec:numericSetup}, we extracted the critical exponent $\nu$ fitting the $\Binder(\PolyIm)$ data for different spatial lattice sizes according to \Eq{eq:BinderFSS}.
Because of the numerical cost the number of simulated $\beta$'s is limited.
If the distributions of $\SGluon(\beta_1)/\beta_1$ and $\SGluon(\beta_2)/\beta_2$ have a good overlap,  
one can use Ferrenberg-Swendsen reweighting~\cite{Ferrenberg:1989ui} to obtain our observable
at $\beta_1<\beta_{new}<\beta_2$.
However, increasing the number of reweighted points can arbitrarily reduce the value of the 
$\chi^2_{\text{NDF}}$ of the fits.
For this reason, we almost always reweighted our data using all simulated $\beta$'s, but 
without adding new points, i.e.~where $\beta_{new}$ is one of the simulated $\beta$.
Exceptions to this are the first-order regions where the the Binder cumulant is very steep and 
a higher resolution in $\beta$ is needed.

Varying the fit interval by range and location there is a multitude
of possible fits with differing results from which the ``good'' ones have to be chosen. Here we outline the criteria of
the filter algorithm used to select our results.
\begin{itemize}
    \item We never extrapolate, i.e.~all fitting intervals are placed such that
          \begin{equation}\label{eq:betaCinInterval}
            \beta_c \in I=\bigl[\beta_{\text{min}},\beta_{\text{max}}\bigr]\;.
          \end{equation}
          
    \item Since the scaling variable is $x\equiv(\beta-\beta_c)N_\sigma^{1/\nu}$,
           the scaling region in $\beta$ shrinks with growing $N_\sigma$.
          Thus, for the fitting intervals $I_1,\dots,I_n$ of the data with 
           $N_{\sigma_1}< \ldots< N_{\sigma_n}$, we demand
          \begin{equation}\label{eq:pyramid}
          I_1\supseteq\dots\supseteq I_n\;.
          \end{equation}

\item On the reduced chi-square we impose
                  \[
              1-\delta\leq\chi^2\leq1+\delta,\qquad \mbox{with}\quad \delta\approx 0.2\;.
              \]
            \item The fitting range in $x$ should ideally be the same for all volumes included.
          We map the intervals $I_n$ to intervals
          \[\tilde{I}_n\equiv\bigl[x_n^{\text{min}},x_n^{\text{max}}\bigr]\;.\]
                  For two intervals $A=[a_1,a_2]$ and $B=[b_1,b_2]$, we define an \emph{overlap percentage} as
          \begin{equation}\label{eq:ovPercentage}
            \Omega\equiv
            \left\{
            \begin{aligned}
                &0 \phantom{\frac{a}{b}}&& \text{if } a_2 < b_1 \vee b_2 < a_1 \\
                &100\cdot\Biggl(1-\frac{\lvert b_1-a_1\rvert+\lvert b_2-a_2\rvert}{a_2-a_1+b_2-b_1}\Biggr) && \text{otherwise}
            \end{aligned}
            \right. \quad .
          \end{equation}
          We then require   $\Omega \geq 80\%$.
                 
    \item Since the scaling region is based on Taylor expansion, it should be symmetric around $x_c$,
                  \[
            I_{\text{scaling}}=\bigl[-\bar{x}, \bar{x}\bigr]\;,
          \]
          with $\bar{x}$ and the size of the region only known after the fit.
          Given an interval $J=[-a,b]$ with $a$ and $b$ non-negative and $a+b$ fixed, we define a \emph{symmetry percentage} as
          \begin{equation}\label{eq:symmPercentage}
            \Xi\equiv 100\cdot\Biggl(1-\biggl\lvert\frac{2a}{a+b}{-1}\biggr\rvert\Biggr) = 100\cdot\Biggl(1-\biggl\lvert\frac{2b}{a+b}{-1}\biggr\rvert\Biggr)  \;.
          \end{equation}
          Clearly, $\Xi=0\%$ (maximally asymmetric interval) for $a=0$ or $b=0$ and $\Xi=100\%$ (maximally symmetric interval) for $a=b$. Among possible fits we choose the one with maximal $\Xi$.
            \end{itemize}

The final list of selected fits is given in Table \ref{tab:fits}.

\vspace{1cm}

\setlength{\tabcolsep}{6pt}
\begin{table}[h]
  \centering
  \begin{tabular}{*{9}{c}c}
    \toprule[0.3mm]
    $\kappa$ & \NSigma & $\beta_c$ & $\nu$ & $\Binder(\beta_c,\infty)$ & $a_1$ & $\chi^2_{_{\text{NDF}}}$ & Q(\%) & $\Omega_{\text{min}}$ & $\Xi_{\text{min}}$ \\[0.5mm]
    \midrule[0.1mm]
    0.1000 & 16 20 24    & 5.86980(29) & 0.43(3)   & 2.141(26) & -0.09(4)  & 1.034 & 41.51 & 86.70 &  6.67 \\
    
    0.1100 & 20 24 30 36 & 5.85670(10) & 0.478(25) & 1.766(11) & -0.14(5)  & 0.999 & 46.26 & 83.06 & 20.00 \\

    0.1200 & 16 20 24 30 & 5.82870(10) & 0.56(3)   & 1.872(8)  & -0.31(10) & 1.005 & 45.61 & 87.18 & 86.00 \\

    0.1300 & 20 24 30    & 5.78670(20) & 0.67(5)   & 1.818(18) & -0.72(28) & 0.980 & 45.82 & 84.12 & 82.50 \\   
    
    \cc{0.1300} & \cc{16 24 32} & \cc{5.95872(26)} & \cc{0.47(1)} & \cc{2.048(8)} & \cc{-0.05(1)} &
    \cc{0.984} & \cc{49.50} & \cc{80.02} & \cc{72.67} \\

    0.1550 & 16 20 24 30 & 5.52840(10) & 0.59(5)   & 1.804(14) & -0.8(3)   & 1.048 & 40.03 & 81.44 & 40.00 \\

    0.1575 & 18 24 30    & 5.48330(10) & 0.648(29) & 1.990(20) & -1.4(3)   & 0.995 & 47.08 & 88.49 & 92.50 \\
    
    0.1600 & 18 24 30    & 5.43670(10) & 0.60(4)   & 1.781(20) & -1.5(5)   & 1.017 & 43.04 & 87.14 & 52.00 \\
    
    0.1625 & 12 18 24    & 5.38620(9)  & 0.471(15) & 1.906(5)  & -0.72(13) & 1.004 & 45.52 & 81.61 & 100.00 \\

    0.1650 & 16 20 24    & 5.33477(3)  & 0.350(20) & 1.680(7)  & -0.15(7)  & 1.007 & 45.40 & 91.40 & 65.00 \\

    \bottomrule[0.3mm]
  \end{tabular}
  \caption{Overview of the selected fits to extract the final value of $\nu$ (the gray background line refers
           to $\NTau=8$). 
           The fits have been performed according to \Eq{eq:BinderFSS}, considering the linear term only.
           The \NSigma{} column contains the spatial lattice extents that have been included in the fits.
           $\Omega_{\text{min}}$ and $\Xi_{\text{min}}$ are respectively the minimum overlap percentage 
           and the minimum symmetry percentage of \Eq{eq:ovPercentage} and \Eq{eq:symmPercentage}.}
  \label{tab:fits}
\end{table}


\begin{thebibliography}{27}%
\makeatletter
\providecommand \@ifxundefined [1]{%
 \@ifx{#1\undefined}
}%
\providecommand \@ifnum [1]{%
 \ifnum #1\expandafter \@firstoftwo
 \else \expandafter \@secondoftwo
 \fi
}%
\providecommand \@ifx [1]{%
 \ifx #1\expandafter \@firstoftwo
 \else \expandafter \@secondoftwo
 \fi
}%
\providecommand \natexlab [1]{#1}%
\providecommand \enquote  [1]{``#1''}%
\providecommand \bibnamefont  [1]{#1}%
\providecommand \bibfnamefont [1]{#1}%
\providecommand \citenamefont [1]{#1}%
\providecommand \href@noop [0]{\@secondoftwo}%
\providecommand \href [0]{\begingroup \@sanitize@url \@href}%
\providecommand \@href[1]{\@@startlink{#1}\@@href}%
\providecommand \@@href[1]{\endgroup#1\@@endlink}%
\providecommand \@sanitize@url [0]{\catcode `\\12\catcode `\$12\catcode
  `\&12\catcode `\#12\catcode `\^12\catcode `\_12\catcode `\%12\relax}%
\providecommand \@@startlink[1]{}%
\providecommand \@@endlink[0]{}%
\providecommand \url  [0]{\begingroup\@sanitize@url \@url }%
\providecommand \@url [1]{\endgroup\@href {#1}{\urlprefix }}%
\providecommand \urlprefix  [0]{URL }%
\providecommand \Eprint [0]{\href }%
\providecommand \doibase [0]{http://dx.doi.org/}%
\providecommand \selectlanguage [0]{\@gobble}%
\providecommand \bibinfo  [0]{\@secondoftwo}%
\providecommand \bibfield  [0]{\@secondoftwo}%
\providecommand \translation [1]{[#1]}%
\providecommand \BibitemOpen [0]{}%
\providecommand \bibitemStop [0]{}%
\providecommand \bibitemNoStop [0]{.\EOS\space}%
\providecommand \EOS [0]{\spacefactor3000\relax}%
\providecommand \BibitemShut  [1]{\csname bibitem#1\endcsname}%
\let\auto@bib@innerbib\@empty
\bibitem [{\citenamefont {Meyer}(2015)}]{Meyer:2015}%
  \BibitemOpen
  \bibfield  {author} {\bibinfo {author} {\bibfnamefont {H.}~\bibnamefont
  {Meyer}},\ }\href@noop {} {\bibfield  {journal} {\bibinfo  {journal} {PoS}\
  }\textbf {\bibinfo {volume} {LAT2015}},\ \bibinfo {pages} {354} (\bibinfo
  {year} {2015})}\BibitemShut {NoStop}%
\bibitem [{\citenamefont {de~Forcrand}\ and\ \citenamefont
  {Philipsen}(2010)}]{deForcrand:2010he}%
  \BibitemOpen
  \bibfield  {author} {\bibinfo {author} {\bibfnamefont {P.}~\bibnamefont
  {de~Forcrand}}\ and\ \bibinfo {author} {\bibfnamefont {O.}~\bibnamefont
  {Philipsen}},\ }\href {\doibase 10.1103/PhysRevLett.105.152001} {\bibfield
  {journal} {\bibinfo  {journal} {Phys.Rev.Lett.}\ }\textbf {\bibinfo {volume}
  {105}},\ \bibinfo {pages} {152001} (\bibinfo {year} {2010})},\ \Eprint
  {http://arxiv.org/abs/1004.3144} {arXiv:1004.3144 [hep-lat]} \BibitemShut
  {NoStop}%
\bibitem [{\citenamefont {D'Elia}\ and\ \citenamefont
  {Sanfilippo}(2009)}]{D'Elia:2009qz}%
  \BibitemOpen
  \bibfield  {author} {\bibinfo {author} {\bibfnamefont {M.}~\bibnamefont
  {D'Elia}}\ and\ \bibinfo {author} {\bibfnamefont {F.}~\bibnamefont
  {Sanfilippo}},\ }\href {\doibase 10.1103/PhysRevD.80.111501} {\bibfield
  {journal} {\bibinfo  {journal} {Phys. Rev.}\ }\textbf {\bibinfo {volume}
  {D80}},\ \bibinfo {pages} {111501} (\bibinfo {year} {2009})},\ \Eprint
  {http://arxiv.org/abs/0909.0254} {arXiv:0909.0254 [hep-lat]} \BibitemShut
  {NoStop}%
\bibitem [{\citenamefont {Bonati}\ \emph {et~al.}(2011)\citenamefont {Bonati},
  \citenamefont {Cossu}, \citenamefont {D'Elia},\ and\ \citenamefont
  {Sanfilippo}}]{Bonati:2010gi}%
  \BibitemOpen
  \bibfield  {author} {\bibinfo {author} {\bibfnamefont {C.}~\bibnamefont
  {Bonati}}, \bibinfo {author} {\bibfnamefont {G.}~\bibnamefont {Cossu}},
  \bibinfo {author} {\bibfnamefont {M.}~\bibnamefont {D'Elia}}, \ and\ \bibinfo
  {author} {\bibfnamefont {F.}~\bibnamefont {Sanfilippo}},\ }\href {\doibase
  10.1103/PhysRevD.83.054505} {\bibfield  {journal} {\bibinfo  {journal}
  {Phys.Rev.}\ }\textbf {\bibinfo {volume} {D83}},\ \bibinfo {pages} {054505}
  (\bibinfo {year} {2011})},\ \Eprint {http://arxiv.org/abs/1011.4515}
  {arXiv:1011.4515 [hep-lat]} \BibitemShut {NoStop}%
\bibitem [{\citenamefont {Philipsen}\ and\ \citenamefont
  {Pinke}(2014)}]{Philipsen:2014rpa}%
  \BibitemOpen
  \bibfield  {author} {\bibinfo {author} {\bibfnamefont {O.}~\bibnamefont
  {Philipsen}}\ and\ \bibinfo {author} {\bibfnamefont {C.}~\bibnamefont
  {Pinke}},\ }\href {\doibase 10.1103/PhysRevD.89.094504} {\bibfield  {journal}
  {\bibinfo  {journal} {Phys. Rev.}\ }\textbf {\bibinfo {volume} {D89}},\
  \bibinfo {pages} {094504} (\bibinfo {year} {2014})},\ \Eprint
  {http://arxiv.org/abs/1402.0838} {arXiv:1402.0838 [hep-lat]} \BibitemShut
  {NoStop}%
\bibitem [{\citenamefont {Alexandru}\ and\ \citenamefont
  {Li}(2013)}]{Alexandru:2013uaa}%
  \BibitemOpen
  \bibfield  {author} {\bibinfo {author} {\bibfnamefont {A.}~\bibnamefont
  {Alexandru}}\ and\ \bibinfo {author} {\bibfnamefont {A.}~\bibnamefont {Li}},\
  }\href@noop {} {\bibfield  {journal} {\bibinfo  {journal} {PoS}\ }\textbf
  {\bibinfo {volume} {LAT2013}},\ \bibinfo {pages} {208} (\bibinfo {year}
  {2013})},\ \Eprint {http://arxiv.org/abs/1312.1201} {arXiv:1312.1201
  [hep-lat]} \BibitemShut {NoStop}%
\bibitem [{\citenamefont {Roberge}\ and\ \citenamefont
  {Weiss}(1986)}]{Roberge:1986mm}%
  \BibitemOpen
  \bibfield  {author} {\bibinfo {author} {\bibfnamefont {A.}~\bibnamefont
  {Roberge}}\ and\ \bibinfo {author} {\bibfnamefont {N.}~\bibnamefont
  {Weiss}},\ }\href {\doibase 10.1016/0550-3213(86)90582-1} {\bibfield
  {journal} {\bibinfo  {journal} {Nucl.Phys.}\ }\textbf {\bibinfo {volume}
  {B275}},\ \bibinfo {pages} {734} (\bibinfo {year} {1986})}\BibitemShut
  {NoStop}%
\bibitem [{\citenamefont {Bonati}\ \emph {et~al.}(2013)\citenamefont {Bonati},
  \citenamefont {D'Elia}, \citenamefont {de~Forcrand}, \citenamefont
  {Philipsen},\ and\ \citenamefont {Sanfillippo}}]{Bonati:2013tqa}%
  \BibitemOpen
  \bibfield  {author} {\bibinfo {author} {\bibfnamefont {C.}~\bibnamefont
  {Bonati}}, \bibinfo {author} {\bibfnamefont {M.}~\bibnamefont {D'Elia}},
  \bibinfo {author} {\bibfnamefont {P.}~\bibnamefont {de~Forcrand}}, \bibinfo
  {author} {\bibfnamefont {O.}~\bibnamefont {Philipsen}}, \ and\ \bibinfo
  {author} {\bibfnamefont {F.}~\bibnamefont {Sanfillippo}},\ }\href@noop {} {\
  (\bibinfo {year} {2013})},\ \Eprint {http://arxiv.org/abs/1311.0473}
  {arXiv:1311.0473 [hep-lat]} \BibitemShut {NoStop}%
\bibitem [{\citenamefont {Philipsen}\ and\ \citenamefont
  {Pinke}(2015)}]{Philipsen:2015eya}%
  \BibitemOpen
  \bibfield  {author} {\bibinfo {author} {\bibfnamefont {O.}~\bibnamefont
  {Philipsen}}\ and\ \bibinfo {author} {\bibfnamefont {C.}~\bibnamefont
  {Pinke}},\ }\href
  {http://inspirehep.net/record/1391160/files/arXiv:1508.07725.pdf} {\bibfield
  {journal} {\bibinfo  {journal} {PoS}\ }\textbf {\bibinfo {volume}
  {LAT2015}},\ \bibinfo {pages} {149} (\bibinfo {year} {2015})},\ \Eprint
  {http://arxiv.org/abs/1508.07725} {arXiv:1508.07725 [hep-lat]} \BibitemShut
  {NoStop}%
\bibitem [{\citenamefont {de~Forcrand}\ \emph {et~al.}(2007)\citenamefont
  {de~Forcrand}, \citenamefont {Kim},\ and\ \citenamefont
  {Philipsen}}]{deForcrand:2007rq}%
  \BibitemOpen
  \bibfield  {author} {\bibinfo {author} {\bibfnamefont {P.}~\bibnamefont
  {de~Forcrand}}, \bibinfo {author} {\bibfnamefont {S.}~\bibnamefont {Kim}}, \
  and\ \bibinfo {author} {\bibfnamefont {O.}~\bibnamefont {Philipsen}},\
  }\href@noop {} {\bibfield  {journal} {\bibinfo  {journal} {PoS}\ }\textbf
  {\bibinfo {volume} {LAT2007}},\ \bibinfo {pages} {178} (\bibinfo {year}
  {2007})},\ \Eprint {http://arxiv.org/abs/0711.0262} {arXiv:0711.0262
  [hep-lat]} \BibitemShut {NoStop}%
\bibitem [{\citenamefont {Fromm}\ \emph {et~al.}(2012)\citenamefont {Fromm},
  \citenamefont {Langelage}, \citenamefont {Lottini},\ and\ \citenamefont
  {Philipsen}}]{Fromm:2011qi}%
  \BibitemOpen
  \bibfield  {author} {\bibinfo {author} {\bibfnamefont {M.}~\bibnamefont
  {Fromm}}, \bibinfo {author} {\bibfnamefont {J.}~\bibnamefont {Langelage}},
  \bibinfo {author} {\bibfnamefont {S.}~\bibnamefont {Lottini}}, \ and\
  \bibinfo {author} {\bibfnamefont {O.}~\bibnamefont {Philipsen}},\ }\href
  {\doibase 10.1007/JHEP01(2012)042} {\bibfield  {journal} {\bibinfo  {journal}
  {JHEP}\ }\textbf {\bibinfo {volume} {1201}},\ \bibinfo {pages} {042}
  (\bibinfo {year} {2012})},\ \Eprint {http://arxiv.org/abs/1111.4953}
  {arXiv:1111.4953 [hep-lat]} \BibitemShut {NoStop}%
\bibitem [{\citenamefont {Binder}(1981)}]{Binder:1981sa}%
  \BibitemOpen
  \bibfield  {author} {\bibinfo {author} {\bibfnamefont {K.}~\bibnamefont
  {Binder}},\ }\href {\doibase 10.1007/BF01293604} {\bibfield  {journal}
  {\bibinfo  {journal} {Z.Phys.}\ }\textbf {\bibinfo {volume} {B43}},\ \bibinfo
  {pages} {119} (\bibinfo {year} {1981})}\BibitemShut {NoStop}%
\bibitem [{\citenamefont {Pelissetto}\ and\ \citenamefont
  {Vicari}(2002)}]{Pelissetto:2000ek}%
  \BibitemOpen
  \bibfield  {author} {\bibinfo {author} {\bibfnamefont {A.}~\bibnamefont
  {Pelissetto}}\ and\ \bibinfo {author} {\bibfnamefont {E.}~\bibnamefont
  {Vicari}},\ }\href {\doibase 10.1016/S0370-1573(02)00219-3} {\bibfield
  {journal} {\bibinfo  {journal} {Phys.Rept.}\ }\textbf {\bibinfo {volume}
  {368}},\ \bibinfo {pages} {549} (\bibinfo {year} {2002})},\ \Eprint
  {http://arxiv.org/abs/cond-mat/0012164} {arXiv:cond-mat/0012164 [cond-mat]}
  \BibitemShut {NoStop}%
\bibitem [{\citenamefont {Duane}\ \emph {et~al.}(1987)\citenamefont {Duane},
  \citenamefont {Kennedy}, \citenamefont {Pendleton},\ and\ \citenamefont
  {Roweth}}]{Duane:1987de}%
  \BibitemOpen
  \bibfield  {author} {\bibinfo {author} {\bibfnamefont {S.}~\bibnamefont
  {Duane}}, \bibinfo {author} {\bibfnamefont {A.~D.}\ \bibnamefont {Kennedy}},
  \bibinfo {author} {\bibfnamefont {B.~J.}\ \bibnamefont {Pendleton}}, \ and\
  \bibinfo {author} {\bibfnamefont {D.}~\bibnamefont {Roweth}},\ }\href
  {\doibase 10.1016/0370-2693(87)91197-X} {\bibfield  {journal} {\bibinfo
  {journal} {Phys. Lett.}\ }\textbf {\bibinfo {volume} {B195}},\ \bibinfo
  {pages} {216} (\bibinfo {year} {1987})}\BibitemShut {NoStop}%
\bibitem [{\citenamefont {Hasenbusch}(2001)}]{Hasenbusch:2001ne}%
  \BibitemOpen
  \bibfield  {author} {\bibinfo {author} {\bibfnamefont {M.}~\bibnamefont
  {Hasenbusch}},\ }\href {\doibase 10.1016/S0370-2693(01)01102-9} {\bibfield
  {journal} {\bibinfo  {journal} {Phys. Lett.}\ }\textbf {\bibinfo {volume}
  {B519}},\ \bibinfo {pages} {177} (\bibinfo {year} {2001})},\ \Eprint
  {http://arxiv.org/abs/hep-lat/0107019} {arXiv:hep-lat/0107019 [hep-lat]}
  \BibitemShut {NoStop}%
\bibitem [{\citenamefont {Joo}\ \emph {et~al.}(2000)\citenamefont {Joo},
  \citenamefont {Pendleton}, \citenamefont {Kennedy}, \citenamefont {Irving},
  \citenamefont {Sexton}, \citenamefont {Pickles},\ and\ \citenamefont
  {Booth}}]{Joo:2000dh}%
  \BibitemOpen
  \bibfield  {author} {\bibinfo {author} {\bibfnamefont {B.}~\bibnamefont
  {Joo}}, \bibinfo {author} {\bibfnamefont {B.}~\bibnamefont {Pendleton}},
  \bibinfo {author} {\bibfnamefont {A.~D.}\ \bibnamefont {Kennedy}}, \bibinfo
  {author} {\bibfnamefont {A.~C.}\ \bibnamefont {Irving}}, \bibinfo {author}
  {\bibfnamefont {J.~C.}\ \bibnamefont {Sexton}}, \bibinfo {author}
  {\bibfnamefont {S.~M.}\ \bibnamefont {Pickles}}, \ and\ \bibinfo {author}
  {\bibfnamefont {S.~P.}\ \bibnamefont {Booth}},\ }\href {\doibase
  10.1103/PhysRevD.62.114501} {\bibfield  {journal} {\bibinfo  {journal} {Phys.
  Rev.}\ }\textbf {\bibinfo {volume} {D62}},\ \bibinfo {pages} {114501}
  (\bibinfo {year} {2000})},\ \Eprint {http://arxiv.org/abs/hep-lat/0005023}
  {arXiv:hep-lat/0005023 [hep-lat]} \BibitemShut {NoStop}%
\bibitem [{\citenamefont {Ferrenberg}\ and\ \citenamefont
  {Swendsen}(1989)}]{Ferrenberg:1989ui}%
  \BibitemOpen
  \bibfield  {author} {\bibinfo {author} {\bibfnamefont {A.~M.}\ \bibnamefont
  {Ferrenberg}}\ and\ \bibinfo {author} {\bibfnamefont {R.~H.}\ \bibnamefont
  {Swendsen}},\ }\href {\doibase 10.1103/PhysRevLett.63.1195} {\bibfield
  {journal} {\bibinfo  {journal} {Phys.Rev.Lett.}\ }\textbf {\bibinfo {volume}
  {63}},\ \bibinfo {pages} {1195} (\bibinfo {year} {1989})}\BibitemShut
  {NoStop}%
\bibitem [{\citenamefont {Borsanyi}\ \emph {et~al.}(2012)\citenamefont
  {Borsanyi} \emph {et~al.}}]{Borsanyi:2012zs}%
  \BibitemOpen
  \bibfield  {author} {\bibinfo {author} {\bibfnamefont {S.}~\bibnamefont
  {Borsanyi}} \emph {et~al.},\ }\href {\doibase 10.1007/JHEP09(2012)010}
  {\bibfield  {journal} {\bibinfo  {journal} {JHEP}\ }\textbf {\bibinfo
  {volume} {09}},\ \bibinfo {pages} {010} (\bibinfo {year} {2012})},\ \Eprint
  {http://arxiv.org/abs/1203.4469} {arXiv:1203.4469 [hep-lat]} \BibitemShut
  {NoStop}%
\bibitem [{\citenamefont {Bach}\ \emph {et~al.}()\citenamefont {Bach},
  \citenamefont {Pinke}, \citenamefont {Sciarra} \emph {et~al.}}]{CL2QCD}%
  \BibitemOpen
  \bibfield  {author} {\bibinfo {author} {\bibfnamefont {M.}~\bibnamefont
  {Bach}}, \bibinfo {author} {\bibfnamefont {C.}~\bibnamefont {Pinke}},
  \bibinfo {author} {\bibfnamefont {A.}~\bibnamefont {Sciarra}},  \emph
  {et~al.},\ }\href@noop {} {\enquote {\bibinfo {title} {{\clqcd}},}\ }\bibinfo
  {howpublished} {\mbox{\url{https://github.com/CL2QCD/cl2qcd}}}\BibitemShut
  {NoStop}%
\bibitem [{\citenamefont {{Khronos Working Group}}()}]{opencl}%
  \BibitemOpen
  \bibfield  {author} {\bibinfo {author} {\bibnamefont {{Khronos Working
  Group}}},\ }\href@noop {} {\enquote {\bibinfo {title} {The {OpenCL}
  {Specification}},}\ }\bibinfo {note}
  {Http://www.khronos.org/registry/cl/}\BibitemShut {NoStop}%
\bibitem [{\citenamefont {Bach}\ \emph {et~al.}(2013)\citenamefont {Bach},
  \citenamefont {Lindenstruth}, \citenamefont {Philipsen},\ and\ \citenamefont
  {Pinke}}]{Bach:2012iw}%
  \BibitemOpen
  \bibfield  {author} {\bibinfo {author} {\bibfnamefont {M.}~\bibnamefont
  {Bach}}, \bibinfo {author} {\bibfnamefont {V.}~\bibnamefont {Lindenstruth}},
  \bibinfo {author} {\bibfnamefont {O.}~\bibnamefont {Philipsen}}, \ and\
  \bibinfo {author} {\bibfnamefont {C.}~\bibnamefont {Pinke}},\ }\href
  {\doibase 10.1016/j.cpc.2013.03.020} {\bibfield  {journal} {\bibinfo
  {journal} {Comput.Phys.Commun.}\ }\textbf {\bibinfo {volume} {184}},\
  \bibinfo {pages} {2042} (\bibinfo {year} {2013})},\ \Eprint
  {http://arxiv.org/abs/1209.5942} {arXiv:1209.5942 [hep-lat]} \BibitemShut
  {NoStop}%
\bibitem [{\citenamefont {Philipsen}\ \emph {et~al.}(2014)\citenamefont
  {Philipsen}, \citenamefont {Pinke}, \citenamefont {Sciarra},\ and\
  \citenamefont {Bach}}]{Philipsen:2014mra}%
  \BibitemOpen
  \bibfield  {author} {\bibinfo {author} {\bibfnamefont {O.}~\bibnamefont
  {Philipsen}}, \bibinfo {author} {\bibfnamefont {C.}~\bibnamefont {Pinke}},
  \bibinfo {author} {\bibfnamefont {A.}~\bibnamefont {Sciarra}}, \ and\
  \bibinfo {author} {\bibfnamefont {M.}~\bibnamefont {Bach}},\ }\href@noop {}
  {\bibfield  {journal} {\bibinfo  {journal} {PoS}\ }\textbf {\bibinfo {volume}
  {LAT2014}},\ \bibinfo {pages} {038} (\bibinfo {year} {2014})},\ \Eprint
  {http://arxiv.org/abs/1411.5219} {arXiv:1411.5219 [hep-lat]} \BibitemShut
  {NoStop}%
\bibitem [{\citenamefont {Bach}\ \emph {et~al.}(2011)\citenamefont {Bach},
  \citenamefont {Kretz}, \citenamefont {Lindenstruth},\ and\ \citenamefont
  {Rohr}}]{Bach2011a}%
  \BibitemOpen
  \bibfield  {author} {\bibinfo {author} {\bibfnamefont {M.}~\bibnamefont
  {Bach}}, \bibinfo {author} {\bibfnamefont {M.}~\bibnamefont {Kretz}},
  \bibinfo {author} {\bibfnamefont {V.}~\bibnamefont {Lindenstruth}}, \ and\
  \bibinfo {author} {\bibfnamefont {D.}~\bibnamefont {Rohr}},\ }\href {\doibase
  10.1007/s00450-011-0161-5} {\bibfield  {journal} {\bibinfo  {journal}
  {Computer Science - Research and Development}\ ,\ \bibinfo {pages} {1}}
  (\bibinfo {year} {2011})}\BibitemShut {NoStop}%
\bibitem [{\citenamefont {Rohr}\ \emph {et~al.}(2015)\citenamefont {Rohr},
  \citenamefont {Bach}, \citenamefont {Neskovic}, \citenamefont {Lindenstruth},
  \citenamefont {Pinke},\ and\ \citenamefont {Philipsen}}]{L-CSC}%
  \BibitemOpen
  \bibfield  {author} {\bibinfo {author} {\bibfnamefont {D.}~\bibnamefont
  {Rohr}}, \bibinfo {author} {\bibfnamefont {M.}~\bibnamefont {Bach}}, \bibinfo
  {author} {\bibfnamefont {G.}~\bibnamefont {Neskovic}}, \bibinfo {author}
  {\bibfnamefont {V.}~\bibnamefont {Lindenstruth}}, \bibinfo {author}
  {\bibfnamefont {C.}~\bibnamefont {Pinke}}, \ and\ \bibinfo {author}
  {\bibfnamefont {O.}~\bibnamefont {Philipsen}},\ }in\ \href@noop {} {\emph
  {\bibinfo {booktitle} {High Performance Computing (LNCS)}}},\ Vol.\ \bibinfo
  {volume} {{9137}}\ (\bibinfo {year} {2015})\BibitemShut {NoStop}%
\bibitem [{\citenamefont {Jin}\ \emph {et~al.}(2015)\citenamefont {Jin},
  \citenamefont {Kuramashi}, \citenamefont {Nakamura}, \citenamefont {Takeda},\
  and\ \citenamefont {Ukawa}}]{Jin:2014hea}%
  \BibitemOpen
  \bibfield  {author} {\bibinfo {author} {\bibfnamefont {X.-Y.}\ \bibnamefont
  {Jin}}, \bibinfo {author} {\bibfnamefont {Y.}~\bibnamefont {Kuramashi}},
  \bibinfo {author} {\bibfnamefont {Y.}~\bibnamefont {Nakamura}}, \bibinfo
  {author} {\bibfnamefont {S.}~\bibnamefont {Takeda}}, \ and\ \bibinfo {author}
  {\bibfnamefont {A.}~\bibnamefont {Ukawa}},\ }\href {\doibase
  10.1103/PhysRevD.91.014508} {\bibfield  {journal} {\bibinfo  {journal} {Phys.
  Rev.}\ }\textbf {\bibinfo {volume} {D91}},\ \bibinfo {pages} {014508}
  (\bibinfo {year} {2015})},\ \Eprint {http://arxiv.org/abs/1411.7461}
  {arXiv:1411.7461 [hep-lat]} \BibitemShut {NoStop}%
\bibitem [{\citenamefont {Philipsen}(2013)}]{Philipsen:2012nu}%
  \BibitemOpen
  \bibfield  {author} {\bibinfo {author} {\bibfnamefont {O.}~\bibnamefont
  {Philipsen}},\ }\href {\doibase 10.1016/j.ppnp.2012.09.003} {\bibfield
  {journal} {\bibinfo  {journal} {Prog. Part. Nucl. Phys.}\ }\textbf {\bibinfo
  {volume} {70}},\ \bibinfo {pages} {55} (\bibinfo {year} {2013})},\ \Eprint
  {http://arxiv.org/abs/1207.5999} {arXiv:1207.5999 [hep-lat]} \BibitemShut
  {NoStop}%
\bibitem [{\citenamefont {Wolff}(2004)}]{Wolff:2003sm}%
  \BibitemOpen
  \bibfield  {author} {\bibinfo {author} {\bibfnamefont {U.}~\bibnamefont
  {Wolff}} (\bibinfo {collaboration} {ALPHA}),\ }\href {\doibase
  10.1016/S0010-4655(03)00467-3, 10.1016/j.cpc.2006.12.001} {\bibfield
  {journal} {\bibinfo  {journal} {Comput. Phys. Commun.}\ }\textbf {\bibinfo
  {volume} {156}},\ \bibinfo {pages} {143} (\bibinfo {year} {2004})},\ \bibinfo
  {note} {[Erratum: Comput. Phys. Commun.176,383(2007)]},\ \Eprint
  {http://arxiv.org/abs/hep-lat/0306017} {arXiv:hep-lat/0306017 [hep-lat]}
  \BibitemShut {NoStop}%
\end{thebibliography}
\end{document}